\documentclass[preprint,showpacs,nofootinbib,amsmath,amssymb,tightenlines]{revtex4}
\usepackage{amssymb,amsbsy,amsmath,amsthm,bm}
\usepackage[mathscr]{eucal}

\theoremstyle{plain}
\newtheorem{theorem}{Theorem}

\newtheorem{lemma}[theorem]{Lemma}
\newtheorem{corollary}[theorem]{Corollary}

\numberwithin{equation}{section}
\numberwithin{theorem}{section}

\newcommand{\startproof}{\setlength{\parindent}{0in}\textbf{Proof.}}
\newcommand{\finishproof}{\hfill $\Box$ \\}

\newcommand{\double}[2]{\hspace{0.1em} #1 \hspace{#2} #1 \hspace{0.1em}}
\newcommand{\ident}{\double{1}{-0.35em}}

\newcommand{\ntensor}[1]{\overset{\scriptstyle #1}{\otimes}}
\newcommand{\ntensorsm}[1]{\overset{\scriptscriptstyle #1}{\otimes}}
\newcommand{\ncart}[1]{\overset{\scriptstyle #1}{\times}}

\newcommand{\mspan}{\mathrm{span}}  
\newcommand{\Ker}{\mathrm{Ker}}

\newcommand{\natiso}{\overset{\scriptscriptstyle \,\mathrm{nat.}}{\cong}}

\newcommand{\C}{\mathbb{C}}
\newcommand{\Q}{\mathbb{Q}}
\newcommand{\R}{\mathbb{R}}
\newcommand{\Z}{\mathbb{Z}}

\newcommand{\rme}{\mathrm{e}}
\newcommand{\rmi}{\mathrm{i}}
\newcommand{\dif}{\mathrm{d}}
\newcommand{\deriv}[2]{\frac{\dif #1}{\dif #2}}
\newcommand{\pderiv}[2]{\frac{\partial #1}{\partial #2}}
\newcommand{\half}{\frac{1}{2}}

\newcommand{\tderiv}[2]{\tfrac{\dif #1}{\dif #2}}
\newcommand{\tpderiv}[2]{\tfrac{\partial #1}{\partial #2}}
\newcommand{\thalf}{\tfrac{1}{2}}

\newcommand{\sshalf}{\genfrac{}{}{}{3}{1}{2}}

\newcommand{\SO}{\mathrm{SO}}
\newcommand{\SU}{\mathrm{SU}}
\newcommand{\Diff}{\mathrm{Diff}}
\newcommand{\Aut}{\mathrm{Aut}}

\newcommand{\sfock}{\mathcal{F}_s}
\newcommand{\Hil}{\mathcal{H}}
\newcommand{\hil}{h}
\newcommand{\Sch}{\mathcal{S}}
\newcommand{\Lie}{\mathcal{L}}
\newcommand{\Cyl}{\mathrm{Cyl}}
\newcommand{\andm}{\mathrm{and}}

\newcommand{\scrD}{\mathcal{D}}
\newcommand{\scrS}{\mathcal{S}}
\newcommand{\scrC}{\mathcal{C}}

\newcommand{\scrP}{\mathscr{P}}

\newcommand{\symgr}{\mathscr{T}}

\newcommand{\dummy}{\rule[0in]{0in}{0in}}

\newcommand{\entry}[2]{
\begin{minipage}[t]{1.5in} #1 \end{minipage}
\hspace{0.25in}
\begin{minipage}[t]{3.75in} \hspace{-0.25in} #2 \end{minipage}
\newline
}

\begin{document}

\preprint{IGPG--05/11--4}
\title{Quantum field theory and its symmetry reduction}
\author{Jonathan Engle}
\email{engle@gravity.psu.edu}
\affiliation{
Institute for Gravitational Physics and Geometry,\\
Physics Department, Penn State, University Park, PA 16802, USA}

\begin{abstract}
The relation between symmetry reduction before and after quantization
of a field theory is discussed using a toy model: the axisymmetric
Klein-Gordon field.  We consider three possible notions of symmetry
at the quantum level: invariance under the group action, and two
notions derived from imposing symmetry as a system of constraints
a la Dirac, reformulated as a first class system.
One of the latter two turns out to be the most appropriate notion
of symmetry in the sense that it satisfies a number of physical
criteria, including the commutativity of quantization and symmetry
reduction.  Somewhat surprisingly, the requirement of
invariance under the symmetry group action is \textit{not}
appropriate for this purpose. A generalization of the physically
selected notion of symmetry to loop quantum gravity is presented and
briefly discussed.
\end{abstract}



\maketitle



\section{Introduction}

In loop quantum cosmology (LQC), an issue of primary importance
is the relation to the full theory, loop quantum
gravity (LQG).  For introductions to loop quantum gravity, see
\cite{lqg_revs}, and for loop quantum cosmology,
see \cite{lqc_revs}.
The issue of the relation to the full theory
is particularly important as possible predictions testable by
cosmological observations are starting to be made based on
LQC \cite{predictions}.  It is important to know to what extent
tests of such possible predictions will in fact be tests of full loop
quantum gravity.

Other symmetry reduced models in loop quantum gravity have also been
constructed, for example, for the purpose of better understanding
quantum black holes \cite{quantumbh_paps}.

In dealing with either of these reduced models, the underlying hope
is that quantization and symmetry reduction
commute in the case of loop quantum gravity, in some sense.
The question of commutation of symmetry
reduction and quantization is an old one.  However, it is sometimes
not appreciated that the question of whether commutation is achieved
depends in a critical way on what one means by the ``symmetric sector''
of the full quantum theory.  One would like the ``symmetric sector''
of the full quantum theory (defined in some physically well-motivated
way) to be isomorphic to
the reduced-then-quantized theory.\footnote{
The phrase ``reduced-then-quantized theory'' is of course just as
ambiguous as the phrase ``quantize.''  What is meant here, roughly,
is the reduced theory quantized as a theory on the reduced spatial
manifold (pp.\pageref{redman1},\pageref{redman2}),
using the same quantization methods as in the full theory.
}
If one can achieve such an isomorphism,
not only at the level of Hilbert space structure, but also at the level
of dynamics, one will have achieved full commutation of symmetry reduction
and quantization.  One may also have partial commutation:
it may be that it is only the Hilbert space structure of the reduced
theory that is isomorphic to the ``symmetric sector'' of the full
theory. Nevertheless, even in such a situation
one can ask if there is some choice of Hamiltonian operator in the
reduced theory that ``best'' represents the information contained in
the Hamiltonian of the full theory.

We will address all of these issues, but in the simple context where
the full theory is well understood:
the axisymmetric, free Klein-Gordon field in Minkowski space.
However, this analysis will suggest a generalization to more interesting
contexts. In particular, in the conclusions, a programme of application
to loop quantum gravity will be sketched.

At first the fact that there is an ``ambiguity'' in
the notion of symmetry seems surprising.  However, there are at least
two possible approaches to defining a notion of symmetry
at the quantum level:
\begin{enumerate}
\item Demanding invariance under the action of the symmetry group
\item 
Taking a system of constraints that classically
isolates the symmetric sector,
and then imposing these constraints as one would
in constrained quantization.
\end{enumerate}
In the case of the axisymmetric Klein-Gordon theory, the notion
of symmetry in sense 1 above is straightforward:
a state is axisymmetric if it is annihilated by $\hat{\mathbb{L}}_z$,
the operator corresponding to the total angular momentum
in the z direction, since this is the generator of the action
of rotations about the z axis.

However, we will also find two distinct, but natural ways of
implementing notion 2, corresponding to two different
ways of reformulating the symmetry constraints as a first
class system --- to be referred to as reformulations A and B, as
defined below.  Thus in this paper we actually consider
three distinct notions of symmetry:
\begin{enumerate}
\item \label{inv_symm}
Requiring invariance under the action of the symmetry group
($\hat{\mathbb{L}}_z \Psi = 0$).
\item \label{A_symm}
Imposition of $\Lie_\phi \hat{\varphi}(x)\Psi = 0$.
(Constraint reformulation A.)
\item \label{B_symm}
Imposition of $a([\Lie_\phi f, \Lie_\phi g])\Psi = 0$.
(Constraint reformulation B.)
\end{enumerate}
where $\phi^a$ is the axial Killing field on Minkowski space
generating rotation about the z axis.
In \ref{B_symm}, $[f,g]$ denotes the phase space point
determined by the initial data $\varphi = f$, $\pi = g$,
with $a([f,g])$ denoting the associated annihilation operator
(see next section).
We shall refer to these notions of symmetry as ``invariance symmetry'',
``A-symmetry'', and ``B-symmetry'', respectively.  A state satisfying
one of these conditions of symmetry will likewise be referred to as
``invariant'', ``A-symmetric'', or ``B-symmetric''.

In the rest of this paper, these three notions of symmetry will be
explained, justified, characterized and compared in
detail.  Simpler, less central results will be stated without proof.
The A-symmetric sector (with appropriate choice of
inner product) and B-symmetric sector of the quantum theory
will turn out to be naturally isomorphic to the reduced-then-quantized
Hilbert space $\Hil_{red}$.  Thus A-symmetry and B-symmetry as notions
of symmetry achieve commutation of quantization and reduction.
Furthermore
\begin{itemize}
\item A-symmetry and B-symmetry are strictly stronger than invariance
  symmetry.  That is, if a state is A-symmetric  or B-symmetric, it is
  also invariant.
\item \label{B_and_Boj}
  The space of A-symmetric states is the space of wavefunctions
  with support only on symmetric configurations. \footnote{
  Although this seems obvious at first, the rigorous
  formulation of this statement is more non-trivial to prove.}
  It is thus the analogue of the notion of symmetry used by
  Bojowald in quantum cosmology.
\item The space of B-symmetric states is equal to the span of the
  set of coherent states associated with the symmetric sector of the
  classical theory.
\item In a precise sense, B-symmetric states are those in which
  all non-symmetric modes are unexcited.
\item For B-symmetric states, fluctuations away from axisymmetry are
  minimized in a precise sense.
\item The quantum Hamiltonian preserves the space of invariant
  states and the space of B-symmetric states, but not the space
  of A-symmetric states. \footnote{
Therefore B-symmetry achieves full commutation of reduction
and quantization, whereas A-symmetry achieves commutation only
at the level of Hilbert space structure.}
\end{itemize}
Because of the last four items on this list, we argue that
B-symmetry should be preferred over A-symmetry as an
embedding of the reduced theory.

We then motivate and discuss a prescription for carrying
arbitrary operators in the full theory over to the reduced theory.
Finally, in the conclusions, we summarize what has been learned and
discuss application to loop quantum gravity.
For convenience of the reader we have collected definitions of
mathematical symbols in appendix \ref{symb_app}.

We conclude with a conceptually important point.  One may object
that commutation of reduction and quantization is achieved only
because we have chosen to use ``indirect'' notions of symmetry, rather
than the obvious notion of invariance under the symmetry group.  But,
in fact, in quantum gravity, if the question of commutation is even to
be \textit{posed} (in a non-trivial way), one \textit{must} use a notion of
symmetry other than invariance symmetry.  For, in quantum gravity,
after the diffeomorphism constraint has been solved, the action of
the symmetry group (if it is a spatial symmetry) is trivial,
and so invariance symmetry becomes a vacuous notion.  The reason for
this is that the symmetry group becomes a subgroup of the
gauge group of the canonical theory.\footnote{
As discussed in the conclusions, this situation can be
exactly mimicked in the axisymmetric Klein-Gordon case by
simply declaring the three components of the total angular
momentum to be constraints, so that the canonical gauge group
is just the group of $\SO(3)$ rotations about the origin.
}
But if this is the case in quantum gravity, perhaps, then, one
should not be surprised if also in other theories
invariance symmetry is inappropriate for commutation questions.
Indeed, in the Klein-Gordon case
at hand, not only is invariance symmetry less desireable
in that it does not achieve commutation, but B-symmetry
satisfies many physical criteria which invariance symmetry does
not (see list of results above).

\section{Preliminaries: review of quantization of the Klein-Gordon field}
\label{sect_prelim}

First, let us review those aspects of the treatment
\cite{qft_refs1} of the quantization of the free Klein-Gordon
theory that will be used in the rest of this paper.
This section will also serve to fix notation.

\subsection{Classical theory}

Let $\Sigma$ denote a fixed Cauchy surface:
a spatial hyperplane in Minkowski space. Let $q_{ab}$ denote
the induced Euclidean metric on $\Sigma$. The phase space $\Gamma$
is a vector space parametrized by two smooth real scalar fields
$\varphi(x)$ and $\pi(x)$ on $\Sigma$ with an appropriate fall-off
at infinity.  (The precise fall-off condition is
unimportant for our purposes.)
The symplectic structure is simply
\begin{equation}
\Omega([\varphi,\pi],[\varphi',\pi'])
= \int_\Sigma (\pi \varphi' - \varphi \pi') \dif^3 x
\end{equation}
so that the Poisson brackets between the basic variables are
\begin{equation}
\label{poissonalg}
\{ \varphi(x), \pi(y) \} = \delta^3(x,y)
\end{equation}
and $\{\varphi(x),\varphi(y)\}=\{\pi(x),\pi(y)\}=0$.
In a word, $\Gamma$ is the cotangent bundle over the space
of all smooth fields $\varphi$ on $\Sigma$ (with appropriate
fall-off imposed).

The Hamiltonian of the scalar field with mass $m$ is
\begin{equation}
\label{hamiltonian}
\mathbb{H} = \half \int_\Sigma \{\pi^2 + (\vec{\nabla}\varphi)\cdot(\vec{\nabla}\varphi)
+ m^2 \varphi^2\} \dif^3 x
\end{equation}
From this Hamiltonian, one derives the equations of motion
to be
\begin{eqnarray}
\label{eom}
\dot{\varphi} &=& \pi
\\
\dot{\pi} &=& \Delta\varphi - m^2 \varphi
\end{eqnarray}
where $\Delta$ is the Laplacian on $\Sigma$.
Let $\Theta:= -\Delta+m^2$.
Choosing the complex structure
\begin{equation}
\label{compl_str}
J [\varphi,\pi] = [-\Theta^{-\half} \pi, \Theta^\half \varphi]
\end{equation}
we turn $\Gamma$ into a complex vector space.
The Hermitian inner product thereby determined on $\Gamma$
is then
\begin{eqnarray}
\nonumber
\langle [\varphi,\pi],[\varphi',\pi'] \rangle
&:=& \half \Omega(J[\varphi,\pi],[\varphi',\pi'])
- \rmi\half\Omega([\varphi,\pi],[\varphi',\pi'])
\\
&=& \half (\Theta^\half \varphi, \varphi')+\half(\Theta^{-\half}\pi,\pi')
-\frac{\rmi}{2}(\pi, \varphi')+\frac{\rmi}{2}(\varphi,\pi') \qquad
\end{eqnarray}
where for $f,g$ functions on $\Sigma$, we define
$(f,g):= \int_\Sigma fg \dif^3 x$.
Completing $\Gamma$ with respect to this Hermitian
inner product gives the single particle Hilbert
space $\hil$.

In constructing the Hilbert space for the field theory,
one then has two possible approaches: the Fock and Schr\"{o}dinger
approaches.

\subsection{Fock quantization}

In the Fock approach, the full Hilbert space is constructed as
\begin{equation}
\Hil := \sfock(\hil) :=
\bigoplus_{n=0}^\infty \left.\bigotimes^n\right._s \,\hil
\end{equation}
where $\ntensor{n}_s \hil$ denotes the
symmetrized tensor product of $n$ copies of $\hil$.
\footnote{
That is, $\ntensor{n} \hil$ is the space of
all continuous multilinear maps $\ncart{n} \hil' \rightarrow \C$;
$\ntensor{n}_s \hil$ is then the space of all members of
$\ntensor{n} \hil$ invariant under arbitrary permutations of arguments.
}

For each $n$, the inner product on $\hil$ induces a unique inner product on
$\ntensor{n} \hil$ via the condition
\begin{equation}
\langle \psi_1 \otimes \psi_2 \otimes \dots \otimes \psi_n,
\phi_1 \otimes \phi_2 \otimes \dots \otimes \phi_n
\rangle_{\ntensorsm{n} \hil}
= \langle \psi_1, \phi_1 \rangle \langle \psi_2, \phi_2 \rangle
\cdots \langle \psi_n, \phi_n \rangle
\end{equation}
for all $\{\psi_i\}, \{\phi_i\} \in \hil$.
This in turn induces an inner product on
$\ntensor{n}_s \hil$.

Let $A,B,C \dots$ denote abstract indicies associated
with the single particle Hilbert space $\hil$.
Let prime denote topological dual.
Then, for each $n$, define the complex conjugation map
$\ntensor{n} \hil \mapsto (\ntensor{n} \hil)'$,
$\psi^{A_1 \dots A_n} \mapsto \overline{\psi}_{A_1 \dots A_n}$
by
\begin{equation}
\overline{\psi}_{A_1 \dots A_n} \phi^{A_1 \dots A_n}
:= \langle \psi, \phi \rangle_{\ntensorsm{n} \hil}.
\end{equation}

A given member $\Psi \in \Hil = \sfock(\hil)$ takes the form
\begin{equation}
\Psi = (\psi, \psi^{A_1}, \psi^{A_1 A_2}, \psi^{A_1 A_2 A_3}, \dots )
\end{equation}
with each component $\psi^{A_1 \dots A_n}$ satisfying
$\psi^{A_1 \dots A_n}=\psi^{(A_1 \dots A_n)}$.
The inner product on $\Hil$ is then defined by
\begin{equation}
\langle \Psi, \Phi \rangle
= \sum_{n=0}^\infty \overline{\psi}_{A_1 \dots A_n}\phi^{A_1 \dots A_n}.
\end{equation}

Given an element $\xi^A = [\varphi,\pi]^A \in \hil$, one has
associated creation and annihilation operators which act on $\Hil$
by
\begin{eqnarray}
a^{\dagger}(\xi)\Psi
&:=& (0,\psi \xi^{A_1}, \sqrt{2} \xi^{(A_1}\psi^{A_2)},
\sqrt{3} \xi^{(A_1}\psi^{A_2 A_3)}, \dots )
\\
a(\xi)\Psi
&:=& (\overline{\xi}_A \psi^A, \sqrt{2}\overline{\xi}_A \psi^{A A_1},
\sqrt{3}\overline{\xi}_A \psi^{A A_1 A_2}, \dots )
\end{eqnarray}
One can check
\begin{equation}
[a(\xi),a^\dagger(\eta)] = \langle \xi, \eta \rangle \ident
\end{equation}
The unique normalized state annihilated by all the
annihilation operators is the vacuum; it is given by
\begin{equation}
\Psi_0 := (1,0,0,0, \dots)
\end{equation}
In terms of the creation and annihilation operators, the
representation of the smeared field operators is given by
\begin{eqnarray}
\label{fieldops_def}
\hat{\varphi}[f] &:=& \rmi\{a([0,f])-a^\dagger([0,f])\}
\\
\hat{\pi}[g] &:=& -\rmi\{a([g,0])-a^\dagger([g,0])\}
\end{eqnarray}
With these definitions, one can check
\begin{equation}
\big[\, \hat{\varphi}[f],\hat{\pi}[g]\, \big] =
\rmi \int_{\Sigma} \dif^3 x f g \equiv \rmi(f,g)
\end{equation}
with all other commutators zero, so that (\ref{fieldops_def}) indeed gives
a representation of the Poisson algebra of smeared field variables
(\ref{poissonalg}).

It is useful to note that, by using the fact that  $a^\dagger(\xi)$ is linear
and $a(\xi)$ is anti-linear in $\xi$, one can invert (\ref{fieldops_def})
to obtain an expression for the creation and annihilation operators
in terms of the field operators:
\begin{eqnarray}
\label{cr_an_interms_of_field}
a([f,g]) &=& \half\hat{\varphi}[\Theta^\half f - \rmi g]
+ \half\hat{\pi}[\Theta^{-\half}g + \rmi f]
\\
\nonumber
a^\dagger([f,g]) &=& \half\hat{\varphi}[\Theta^\half f + \rmi g]
+ \half\hat{\pi}[\Theta^{-\half}g - \rmi f]
\end{eqnarray}
These expressions can then be carried over to the classical theory
to obtain functions on $\Gamma$ which are classical analogues
of the creation and annihilation operators.  Upon simplifying
the expressions for these classical analogues, one obtains the remarkably
simple result
\begin{eqnarray}
\label{annih_op_anal}
a([f,g]) &=& \langle [f,g], [\varphi,\pi] \rangle
\\
\label{cr_op_anal}
a^\dagger([f,g]) &=& \langle [\varphi,\pi], [f,g] \rangle
\end{eqnarray}
The Poisson brackets among these classical analogues
exactly mimic the commutators of the quantum
counterparts.

Next, we quantize the Hamiltonian.  Rewriting the classical Hamiltonian
(\ref{hamiltonian}),
\begin{eqnarray}
\mathbb{H} &=& \half \int_\Sigma (\pi^2 - \varphi \Delta \varphi + m^2 \varphi^2 )\dif^3 x
\\
&=& \half \int_\Sigma (\pi^2 + \varphi \Theta \varphi )\dif^3 x
\end{eqnarray}
From (\ref{eom}) and (\ref{compl_str}), we obtain the single particle Hamiltonian
operator on $\hil$,
\begin{eqnarray}
\hat{H}[\varphi, \pi] &:=& J \deriv{}{t}[\varphi,\pi]
\\
&=& [\Theta^\half \varphi, \Theta^\half \pi]
\end{eqnarray}
In terms of this, the classical Hamiltonian can be expressed as
\begin{equation}
\mathbb{H} = \langle [\varphi,\pi], \hat{H} [\varphi,\pi] \rangle
\end{equation}
Let $\{\xi_i = [f_i,g_i]\}$ denote an arbitrary orthonormal
basis of $\hil$.  Then
\begin{eqnarray}
\mathbb{H} &=& \sum_{i,j}\langle [\varphi,\pi],\xi_i\rangle \langle \xi_i, \hat{H} \xi_j \rangle
\langle \xi_j, [\varphi,\pi]\rangle
\\
&=& \sum_{i,j}\langle \xi_i, \hat{H} \xi_j \rangle a^\dagger(\xi_i) a(\xi_j)
\end{eqnarray}
Which is an expression that can be taken directly over to the
quantum theory, using normal ordering:
\begin{equation}
\label{fockham}
\hat{\mathbb{H}} =
\sum_{i,j}\langle \xi_i, \hat{H} \xi_j \rangle a^\dagger(\xi_i) a(\xi_j)
\end{equation}

\subsection{Schr\"{o}dinger quantization}
\label{schr_quant}

As mentioned, the classical phase space $\Gamma$ has a cotangent bundle structure $T^* \scrC$
over some appropriately defined configuration space $\scrC$.

In the case of a finite number of degrees of freedom,
the standard way to quantize a cotangent bundle $T^* \scrC$ is via
a Schr\"{o}dinger representation -- that is, a representation
of the field operators on an $L^2(\scrC,\dif \mu)$ for some
appropriately chosen measure $\mu$.

In the field theory case, however, the measures one is interested
in using are usually not supported on the classical configuration space
$\scrC$, but rather on some appropriate distributional-like
extension $\overline{\scrC}$.  This extension is referred to as
the \textit{quantum configuration space}.

In the case of the free Klein-Gordon field in Minkowski space,
the appropriate quantum configuration space can be taken to be the
space of tempered distributions $\scrS'(\Sigma)$ on $\Sigma$ \cite{gj}.
$\scrS(\Sigma)$ denotes the space of Schwarz functions equipped
with the appropriate topology \cite{rs}, and the prime indicates
the topological dual.  From here on $\varphi$ will denote an element
of $\Sch'(\Sigma)$.

The appropriate measure is the Gaussian measure heuristically given
by the expression
\begin{equation}
\label{heurist_meas}
\text{``}\dif \mu = \exp\left\{-\half(\varphi, \Theta^\half \varphi)\right\}
\scrD \varphi \text{''}
\end{equation}
where $\scrD \varphi$ is the fictitious translation-invariant ``Lesbesgue'' measure on $\scrS'(\Sigma)$.
To more rigorously define the measure, one can specify its
\textit{Fourier transform}.  The Fourier transform of a measure
$\mu$ is defined by
\begin{equation}
\chi_\mu(f):= \int_{\varphi \in \scrS'(\Sigma)}\rme^{\rmi\varphi(f)}\dif \mu
\end{equation}
for $f \in \scrS(\Sigma)$.
The Fourier transform giving rise to
(the rigorous version of) the measure in
(\ref{heurist_meas}) is
\begin{equation}
\chi_\mu(f)=\exp\left\{-\half(f,\Theta^{-\half}f)\right\}
\end{equation}
For further details, see \cite{gj}.

$\Hil = L^2(\scrS'(\Sigma),\dif \mu)$ is then the Hilbert space
of states in the quantum field theory.
For this paper it will also be necessary to introduce a certain
dense subset of $\Hil$ --- the space of \textit{cylindrical functions}.
A function $\Psi:\scrS'(\Sigma)\rightarrow \C$ is
called \textit{cylindrical}
if $\Psi[\varphi] = F(\varphi(e_1),\dots,\varphi(e_n))$
for some $\{e_1, \dots, e_n\} \subseteq \scrS(\Sigma)$
(referred to as ``probes'')
and some smooth function $F:\R^n \rightarrow \C$ (with growth
less than exponential).
More specifically, such a $\Psi$
is said to be \textit{cylindrical with respect to}
the ``probes'' $e_1,\dots,e_n$.
Let the space of cylindrical functions be
denoted $\Cyl$.  

Next, the representation of the field observables on $\Hil$ is
\begin{eqnarray}
\label{schro_config_op}
(\hat{\varphi}[f] \Psi)[\varphi] &:=& \varphi[f] \Psi[\varphi]
\\
\nonumber
(\hat{\pi}[g] \Psi)[\varphi] &:=&
\left[
\text{Self-adjoint part of } -\rmi\int_\Sigma \dif^3 x g \frac{\delta}{\delta \varphi}\right] \Psi[\varphi]
\\
\label{schro_moment_op}
&=& -\rmi \int_\Sigma \dif^3 x \left(g \frac{\delta}{\delta \varphi} - \varphi \Theta^\half g \right)
\Psi[\varphi]
\end{eqnarray}

We then use equations (\ref{cr_an_interms_of_field})
to define creation and annihilation operators in the Schr\"{o}dinger picture.
Substituting (\ref{schro_config_op}) and (\ref{schro_moment_op}) into these
expressions and simplifying, we obtain
\begin{eqnarray}
\label{schr_an_op}
a([f,g]) &=& \half \int_\Sigma \dif^3 x \left(f - \rmi\Theta^{-\half}g\right)
\frac{\delta}{\delta \varphi}
\\
\label{schr_cr_op}
a^\dagger([f,g]) &=& \hat{\varphi}[\Theta^\half f + \rmi g]
- \half \int_\Sigma \dif^3 x \left(f + \rmi\Theta^{-\half}g\right)
\frac{\delta}{\delta \varphi}
\end{eqnarray}
(In (\ref{schr_an_op}), the $\hat{\varphi}$ terms exactly cancel,
leaving only a $\delta/\delta \varphi$ term.)
The unique normalized state in the kernel of all of the
annihilation operators is
\begin{equation}
\Psi_0[\varphi] \equiv 1
\end{equation}
The availability of a vacuum state and creation and annihilation
operators in the Schr\"{o}dinger picture allows one to construct
a mapping from the Fock Hilbert space into the Schr\"{o}dinger
Hilbert space. One finds that the mapping is unitary,
so that the Fock and Schr\"{o}dinger descriptions
of the theory are equivalent.

Substituting (\ref{schr_an_op}),(\ref{schr_cr_op}) into
(\ref{fockham}) and simplifying,
\begin{equation}
\hat{\mathbb{H}} =
\int_{\Sigma^2} \dif^3 x \dif^3 y A(x,y) \varphi(y)
\frac{\delta}{\delta \varphi(x)}
- \int_{\Sigma^2} \dif^3 x \dif^3 y B(x,y)
\frac{\delta^2}{\delta \varphi(x) \delta \varphi(y)}
\end{equation}
where
\begin{eqnarray}
A(x,y) &=& \half \sum_{i,j}\langle\xi_i,\hat{H}\xi_j\rangle
(f_j - \rmi \Theta^{-\half}g_j)(x)(\Theta^\half f_i + \rmi g_i)(y)
\\
B(x,y) &=& \frac{1}{4} \sum_{i,j}\langle\xi_i,\hat{H}\xi_j\rangle
(f_i + \rmi \Theta^{-\half}g_i)(x) (f_j - \rmi \Theta^{-\half}g_j)(y)
\end{eqnarray}
where, as before, $\{\xi_i \equiv [f_i, g_i]\}$ is
an orthonormal basis of the single particle Hilbert space.
By integrating $A(x,y)$ and $B(x,y)$ against
test functions, one can show that $A(x,y)$ is the integral
kernel of $\Theta^\half$ and $B(x,y)=\half\delta^3(x,y)$.
Thus,
\footnote{
To our knowledge, expression (\ref{schr_ham})
has not appeared in the literature.
}
\begin{equation}
\label{schr_ham}
\hat{\mathbb{H}} = \int_\Sigma \dif^3 x \left\{
(\Theta^\half \varphi)(x) \frac{\delta}{\delta \varphi(x)}
-\half \frac{\delta^2}{\delta \varphi(x) \delta \varphi(x)} \right\}
\end{equation}
At first the rigorous meaning of this expression may not be obvious.
However, note that for any cylindrical function
$\Psi[\varphi] = F(\varphi(e_1),\dots,\varphi(e_n))$,
\begin{equation}
\frac{\delta}{\delta \varphi(x)} \Psi[\varphi]
= \sum_{i=1}^n e_i(x) (\partial_i F)(\varphi(e_1),\dots,\varphi(e_n)).
\end{equation}
Therefore, the action of $\hat{\mathbb{H}}$ (\ref{schr_ham}) on
the space of cylindrical functions, $\Cyl$, is well defined.
Furthermore one can check that $\hat{\mathbb{H}}$ preserves $\Cyl$.
In proving this, the fact that there is no term quadratic in
$\varphi$ in (\ref{schr_ham}) is important.\footnote{
The reason the $\varphi^2$ term is absent is
that our quantum measure is Gaussian.
Thus there is a tight relation between kinematics (the choice of
measure, and hence the representation of the quantum algebra)
and dynamics (the Hamiltonian operator).  This relates to
the usual statement that in quantum field theory ``dynamics
dictates the choice of kinematics''!
}
Since $\Cyl$ is dense in $\Hil$, we may take $\Cyl$ to be the domain
of $\hat{\mathbb{H}}$; with this domain choice, one can show that
$\hat{\mathbb{H}}$ is essentially self-adjoint.  Thus,
$\hat{\mathbb{H}}$ with domain $\Cyl$ has a unique self-adjoint
extension, and it is this self-adjoint extension that we henceforth
take to be the meaning of $\hat{\mathbb{H}}$.

\section{Some different methods of imposing symmetry}
\label{sect_diffmeth}

\subsection{Classical analysis}
\label{sect_classanal}

As mentioned in the introduction, incorporation of symmetry
by requiring invariance under the action of the symmetry group is
straightforward in the present context: it corresponds
to requiring a state to be annihilated by the operator
$\hat{\mathbb{L}}_z$ corresponding to the z component of the
total angular momentum.
However, selection of the symmetric sector via imposition of
a system of constraints deserves further explanation.

Classically the condition for symmetry takes the form of the constraints
\begin{equation}
\Lie_\phi \varphi = 0 \quad \andm \quad \Lie_\phi \pi = 0
\end{equation}
If we smear the constraints, they take the form
\begin{equation}
\label{fullconstraints}
\varphi[\Lie_\phi f] = 0 \quad \andm \quad \pi[\Lie_\phi f] = 0
\end{equation}
for all test functions $f$ in $\scrS(\Sigma)$,
the space of Schwarz functions.
The form of the smearings $\Lie_{\phi} f$ and $\Lie_{\phi} g$ comes
from an integration by parts.
More generally, the significance of the form $\Lie_\phi f$ for test functions
is the following.  Let $\Sch(\Sigma)_{inv}$ denote the space of elements of
$\Sch(\Sigma)$ Lie dragged by $\phi^a$.  One can show that the space all test
functions of the form $\Lie_\phi f$ is precisely the orthogonal complement of
$\Sch(\Sigma)_{inv}$ in $\Sch(\Sigma)$ (with respect to the usual inner product).
Thus, another way to view the above set of smeared constraints is
that they are the non-symmetric components of the fields;
by requiring these to vanish, we impose symmetry.

Therefore, what we would ideally like to do in the quantum theory is
impose
\begin{equation}
\label{secclass_constr}
\hat{\varphi}[\Lie_\phi f]\Psi = 0 \quad \andm \quad
\hat{\pi}[\Lie_\phi f]\Psi = 0
\end{equation}
for all $f \in \scrS(\Sigma)$.  However, the proposed system of
constraints is second class, and, as Dirac taught us, such systems of
constraints cannot be consistently imposed in quantum theory in this
fashion.  One will find that the unique solution to these constraints
is the zero vector.

To get around this difficulty, the strategy is to reformulate the
constraints (\ref{fullconstraints}) as an equivalent first class system.
We consider two such reformulations:
\begin{enumerate}
\item[(A)] the set of constraints
$\{\varphi[\Lie_\phi f]\}_{f\in \scrS(\Sigma)}$
\item[(B)] the set of constraints
$\{a([\Lie_\phi f, \Lie_\phi g])\}_{f,g \in \scrS(\Sigma)}$
\end{enumerate}
We will refer to these as constraint set (A) and constraint set (B).
Note that $a([f,g])$ here is the classical analogue of
the annihilation operator as given in (\ref{annih_op_anal}).
Thus, constraint set (B) consists in complex linear combinations
of the constraints in (\ref{fullconstraints}).
Each of the constraint sets (A) and (B) forms a first class system.
Although the constraint set (A) is obtained by simply
dropping all the constraints on momenta, nevertheless as explained
below (A) is in a certain sense (relevant for quantum theory)
equivalent to the full set of constraints.

We should also mention that other proposals for imposing second
class constraints have been made in the past, such as that proposed
in Klauder's `universal procedure' for imposing constraints
\cite{klauder}.  There is in fact a relation between approach (B)
here and Klauder's approach: The former is a case of the latter with
some natural choices made. This is discussed later on in section
\ref{sect_tensref} of this paper. In addition approach (B) has
similarities to the method of imposing second class constraints
discussed in \cite{tate}, as was noticed after this work was
completed.

Let us introduce some notation.
Let $\Gamma = \{[\varphi,\pi]\}$ be the full classical phase space.
Let
\begin{enumerate}
\item[]
$\Gamma_{inv}:= \{[\varphi,\pi]\in\Gamma \mid \Lie_\phi \varphi = 0
\; \andm \; \Lie_\phi\pi = 0\}$
\item[]
$\Gamma_A:= \{[\varphi,\pi]\in\Gamma \mid \Lie_\phi \varphi = 0 \}$
\item[]
$\Gamma_B:= \{[\varphi,\pi]\in\Gamma \mid
a([\Lie_\phi f, \Lie_\phi g])\mid_{[\varphi,\pi]}=0
\quad \forall f,g \in \Sch(\Sigma)\}$
\end{enumerate}
So that $\Gamma_A$ is the constraint surface associated with constraint
set (A), and $\Gamma_B$ is the constraint surface associated with
constraint set (B).

\textbf{Analysis of constraint set A}

Since constraint set (A) is obtained by dropping constraints from the full set
(\ref{fullconstraints}), it is not surprising that
$\Gamma_A$ is larger than $\Gamma_{inv}$.  However, the
symplectic structure induced on $\Gamma_A$ via pull-back,
$\Omega_A := i^* \Omega$, is degenerate -- as we should expect since
constraint set (A) is first class.
The degenerate directions
are just the ``gauge'' generated by the constraints $\varphi[\Lie_\phi f]$,
namely $\pi(x) \mapsto \pi(x) + \Lie_\phi f$.
If we divide out by this
``gauge,'' the resulting manifold, $\hat{\Gamma}_A$ is naturally
isomorphic to $\Gamma_{inv}$.

One may object: this notion of ``gauge'' is not \textit{physical} gauge;
it is gauge generated by constraints that we have imposed completely
by hand.  This is true, but the point is that
when a constraint is imposed at the quantum level, you automatically
divide out by the corresponding ``gauge'' \textit{whether or not the gauge
is ``physical''}.

At the quantum level, we will find that the solution to constraint set (A),
when equipped with an appropriate inner product, is naturally isomorphic to
the Hilbert space one obtains when first reducing and then quantizing.
The fact that $\hat{\Gamma}_A$ is naturally isomorphic
to $\Gamma_{inv}$ is the imprint of this fact on the classical theory.

A final important note about constraint set (A) is that
its elements \textit{do not} weakly Poisson-commute with the
total Hamiltonian for the free scalar field (\ref{hamiltonian}).
This foreshadows the fact that in the quantum theory,
the total Hamiltonian operator
will not preserve the solution space to constraint set (A).

\textbf{Analysis of constraint set B}

First, it is important to note that the classical observable
$a([f,g])$, when expanded out as
$a([f,g])= \langle [f,g],[\varphi,\pi] \rangle$,
is a complex linear combination of the constraints (\ref{fullconstraints}).
In fact, in rewriting the full constraint set (\ref{fullconstraints})
as constraint set (B), \textit{no constraints have been dropped}.
Rather, one has reduced the number of constraints by half by
simply taking complex linear combinations of the original constraints.

It is easy to see how this works in a simpler example.
Suppose we are working in a theory in which $\{x_1,x_2,x_3,p_1,p_2,p_3\}$
are the basic variables, and we want to impose the second class system
of constraints $x_3=0$, $p_3=0$.  The analogue of reformulation (A)
in this context would be to just drop the $p_3=0$ constraint.  The
analogue of reformulation (B) would be to replace the two constraints
with the single constraint $z_3:=x_3+ip_3=0$.  Obviously $z_3$, being
only a single constraint, makes up a first class system of constraints.
Nevertheless, classically, $z_3=0$ is completely equivalent to
$x_3=0$ and $p_3=0$.  This is one of the strengths of reformulation
strategy (B): the reformulation is classically completely equivalent
to the original set of constraints, but is now a first class system
so that it can be imposed consistently in quantum theory.

But one may object: how is this possible?  You cannot change the fact
that a certain constraint submanifold is first or second class merely
by reformulating it in terms of different constraints because
first-class and second-class character are \textit{geometrical}
properties of the constraint submanifold \cite{ashtekar1}.
This is indeed true.  Our underlying constraint
submanifold is still \textit{geometrically} a second-class constraint
surface. We have merely allowed it to be \textit{formally} expressed as a
first class system by allowing our constraints to be complex.
But fortunately, for a system of constraints to be consistently
implementable in quantum theory, it is sufficient that they
be only formally first class -- i.e., that their Poisson brackets
with each other vanish weakly.

So, $\Gamma_B = \Gamma_{inv}$.

Another fact that is important to note is that all the elements
of constraint set (B) weakly Poisson-commute with the full Hamiltonian $\mathbb{H}$.
This points to the fact that, in quantum theory, the full Hamiltonian operator
$\hat{\mathbb{H}}$ \textit{will} preserve the solution space to
constraint set (B).

\subsection{Setting up the quantum analysis}

Recall that $\Cyl$ denotes the space of cylindrical functions
on $\Sch'(\Sigma)$.
Let $\Cyl^*$ denote its algebraic dual.

Let
\begin{subequations}
\begin{eqnarray}
\Hil_{inv}&:=& \{\Psi\in\Hil \mid \hat{\mathbb{L}}_z \Psi = 0 \}
\\
\Cyl^*_{inv}&:=& \{\eta\in\Cyl^* \mid \hat{\mathbb{L}}^*_z \eta = 0 \}
\\
\label{cyla_def}
\Cyl^*_A &:=& \{\eta\in\Cyl^* \mid \hat{\varphi}[\Lie_\phi f]^* \eta = 0
\quad \forall f\in \scrS(\Sigma)\}
\\
\Hil_B &:=& \{\Psi\in\Hil \mid a([\Lie_\phi f,\Lie_\phi g])\Psi =0
\quad \forall f,g\in\scrS(\Sigma)\}
\end{eqnarray}
\end{subequations}
$\Hil_{inv}$ and $\Cyl^*_{inv}$ are the sets of elements in
$\Hil$ and $\Cyl^*$ fixed by the natural action of rotations about the
z-axis, whence they are implementations of ``invariance symmetry,''
the first notion of symmetry mentioned in the introduction.
($\Cyl^*_{inv}$ has been introduced simply for the purpose
of comparison with $\Cyl^*_A$.)

$\Cyl^*_A$ is the solution space for constraint set (A) at the quantum
mechanical level.  Constraint set (A) forces its solutions to have
support only on symmetric configurations, as we shall see below.
The space of symmetric configurations has measure zero with respect
to the quantum measure $\mu$ on $\scrS'(\Sigma)$.  Since $\mu$
characterizes the inner product in $\Hil$, all solutions to (A) in
$\Hil$ thus have norm zero, whence one must go to $\Cyl^*$ to find
non-trivial solutions.  In other words, constraint set (A) admits
only non-normalizable solutions.

In addition, one should note that the characterization of
$\Cyl^*_A$ as the space
of functions with support only on symmetric configurations
makes $\Cyl^*_A$ the analogue of the notion of symmetry
used by Bojowald to embed loop quantum cosmology
and other symmetry reduced models into full loop quantum gravity
\cite{symmstates_paps}.

$\Hil_B$ is the solution space for constraint set (B) at the quantum
mechanical level.

\section{The structure of $\Hil$ as $\Hil_{red} \otimes \Hil_{\perp}$}
\label{sect_struct}

Before entering further into a quantum analysis of the different notions of
symmetry, it will be convenient to develop apparatus for relating
the Hilbert space in the full theory ($\Hil$) to the Hilbert space in the
reduced theory ($\Hil_{red}$).
The reduced theory is derived in appendix \ref{redtheory}.

We will denote the group of rotations about the z-axis
by $\symgr \subset \Diff(\Sigma)$.
In the reduced theory,
the spatial manifold is taken to be \label{redman1}
$B:=\Sigma/\symgr$, and the
quantum configuration space $\scrS'(B)$.
Let $P:\Sigma \rightarrow B$ denote canonical projection.
Let $\scrS'(\Sigma)_{inv}$ and $\scrS(\Sigma)_{inv}$
denote the $\symgr$-invariant subspaces of $\scrS'(\Sigma)$
and $\scrS(\Sigma)$, respectively. $\scrS(\Sigma)_{inv}$ is then
naturally identifiable with $\scrS(B)$; we make this
identification. Define
$I: \scrS'(\Sigma)_{inv} \rightarrow \scrS'(B)$
by $[I(\alpha)](f):= \alpha(P^* f)$.
Let $\pi: \scrS(\Sigma) \rightarrow \scrS(B)$ denote group averaging
with respect to the action of $\symgr$.  We here use ``group
averaging'' in a more general sense than usual in that we are not
group averaging ``states.''  It will be convenient in this paper to
let ``group averaging'' have this more general meaning of averaging
elements of any vector space over the action of a group.
One can show the pull-back
$\pi^*: \scrS'(B) \rightarrow \scrS'(\Sigma)$ is the inverse
of $I$, so that
\begin{lemma}
$I$ is an isomorphism.
\end{lemma}
\noindent
Thus $\scrS'(\Sigma)_{inv}$ and $\scrS'(B)$ are naturally isomorphic.
Because of this, henceforth we will simply
identify these two spaces.  That is, the isomorphism
$I$ will sometimes not
be explicitly written.  In addition, we will sometimes
implicitly use the fact that $I$ is compatible
with the structure of the cylindrical functions.
Let $\Cyl_{red}$ denote the space of cylindrical functions
in the reduced theory.  We then have
\begin{lemma}
If $\Phi \in \Cyl$, then
$\Phi \circ I^{-1} \in \Cyl_{red} \subseteq \Hil_{red}$,
and the map $\Phi \mapsto \Phi \circ I^{-1}$ is onto $\Cyl_{red}$.
\end{lemma}
\noindent
Next, let $\Pi: \Sch'(\Sigma) \rightarrow \Sch'(\Sigma)_{inv}$
denote group averaging on $\Sch'(\Sigma)$ with respect to $\symgr$.
Recall the quantum measure $\mu$ on $\Sch'(\Sigma)$ introduced
in section \ref{schr_quant}. We then have the following result, which
will be important in theorem \ref{meas_sep}:
\begin{lemma}
\label{consistentmeas}
$\Pi_* \mu = \mu_{red}$,
where $\mu_{red}$ is the quantum measure in the reduced
theory as constructed in appendix \ref{redtheory}.
\end{lemma}
\noindent
Let $\scrS'(\Sigma)_{\perp} := \mathrm{Ker} \,\Pi$.
\begin{lemma}
$\scrS'(\Sigma) = \scrS'(\Sigma)_{inv} \oplus \scrS'(\Sigma)_{\perp}$.
\end{lemma}
\noindent
Topologically, then, $\scrS'(\Sigma)$ has the structure
$\scrS'(\Sigma) = \scrS'(\Sigma)_{inv} \times \scrS'(\Sigma)_{\perp}$.
In terms of this structure, $\Pi$ is canonical projection
into the first factor. We will adopt the convention that if
$\varphi \in \scrS'(\Sigma)$, then $\varphi_s$ and
$\varphi_{\perp}$ denote the components of $\varphi$
with respect to the above decomposition.
\begin{theorem}
\label{meas_sep}
$\mu$ is separable over
$\scrS'(\Sigma) = \scrS'(\Sigma)_{inv} \times \scrS'(\Sigma)_{\perp}$.
That is, there exists a measure $\mu_{inv}$ on $\scrS'(\Sigma)_{inv}$
and $\mu_{\perp}$ on $\scrS'(\Sigma)_{\perp}$, unique up to rescaling,
such that $\mu = \mu_{inv} \times \mu_{\perp}$.  If we furthermore
require $\Pi_* \mu = \mu_{inv}$, this rescaling freedom is fixed, in
which case, by Lemma \ref{consistentmeas} above,
$\mu_{inv}$ is precisely $\mu_{red}$.
\end{theorem}
\noindent
It follows that
\begin{equation}
L^2(\scrS'(\Sigma),\dif \mu) = L^2(\scrS'(\Sigma)_{inv},\dif \mu_{red})
\otimes L^2(\scrS'(\Sigma)_{\perp}, \dif \mu_{\perp})
\end{equation}
so that if we define
$\Hil_{\perp} := L^2(\scrS'(\Sigma)_{\perp}, \dif \mu_{\perp})$,
\begin{equation}
\Hil = \Hil_{red} \otimes \Hil_{\perp}.
\end{equation}
\begin{theorem}
\label{separable_ham}
In terms of this tensor product structure of $\Hil$,
the Hamiltonian operator for the free scalar field theory takes the form
\begin{equation}
\label{sep_ham_eq}
\hat{\mathbb{H}} = \hat{\mathbb{H}}_{red} \otimes \ident
+ \ident \otimes \hat{\mathbb{H}}_{\perp}
\end{equation}
where
\begin{equation}
\label{ham_sympart}
\hat{\mathbb{H}}_{red}
= \int \dif^3 x \left\{
(\Theta^{\half} \varphi_s)(x)\frac{\delta}{\delta \varphi_s(x)}
- \half \frac{\delta^2}{\delta \varphi_s (x)^2} \right\}
\end{equation}
is the Hamiltonian of the reduced theory (see appendix \ref{redtheory})
and
\footnote{
In this paper, $\frac{\delta}{\delta \varphi}$, $\frac{\delta}{\delta \varphi_s}$,
and $\frac{\delta}{\delta \varphi_{\perp}}$ are defined with respect to the
volume form $\dif^3 x := \rho \dif \rho \dif z \dif \phi$.  In the
reduced theory, in appendix (\ref{redtheory}), $\frac{\delta}{\delta\varphi_r}$ is
defined with respect to $\dif^2 x := \dif \rho \dif z$.
}
\begin{equation}
\label{ham_perppart}
\hat{\mathbb{H}}_{\perp}
= \int \dif^3 x \left\{
(\Theta^{\half} \varphi_{\perp})(x)\frac{\delta}{\delta \varphi_{\perp}(x)}
- \half \frac{\delta^2}{\delta \varphi_{\perp} (x)^2} \right\}
\end{equation}
\end{theorem}
That is, the Hamiltonian is separable over the tensor product decomposition
of $\Hil$. With these structures established, we
proceed with an analysis of $\Cyl_A^*$ and $\Hil_B$.

\section{Analysis of $\Cyl^*_A$}
\label{cyla_anal}

It will be useful to first prove
the precise way in which
$\Cyl^*_A$ is the space of all elements of $\Cyl^*$ with support
in $\scrS'(\Sigma)_{inv}$.

Define $\Cyl_\sim := \{\Psi \in \Cyl \mid
\mathrm{Supp}\, \Psi \cap \scrS'(\Sigma)_{inv} = \emptyset \}$.
Then we say $\eta \in \Cyl^*$ has support
on $\scrS'(\Sigma)_{inv}$ if $\eta$ is zero on $\Cyl_\sim$.
\begin{lemma}
For $\Psi \in \Cyl$, $\Psi \in \Cyl_\sim$ iff
$\Psi$ is of the form
$\sum_{i=1}^n \varphi(\Lie_\phi f_i) \Phi_i$
for some
$\{f_i\} \subset \Sch(\Sigma)$ and some $\{\Phi_i\} \subset \Cyl$.
\end{lemma}
{\startproof

($\Leftarrow$) obvious.

($\Rightarrow$) Suppose $\Psi \in \Cyl_\sim$. As an element of $\Cyl$,
$\Psi[\varphi]$ depends on $\varphi$ only via a finite number of ``probes''
(see section \ref{sect_prelim}).  There is an ambiguity in how one
chooses the probes; what is important is the finite dimensional
subspace of $\Sch(\Sigma)$ spanned by these probes.  Let $V$ denote
this finite dimensional subspace.  We may then choose any
set of probes spanning $V$ to represent $\Psi$ as a cylindrical function.
Using the decomposition
$\Sch(\Sigma) = \Sch(\Sigma)_{inv} \oplus \Sch(\Sigma)_{\perp}$
(see section \ref{sect_struct}), we demand that our choice of
probes spanning $V$ be a set of the form
$\{\Lie_\phi f_1, \dots \Lie_\phi f_n,e_1,\dots,e_m\}$
where $\Lie_\phi f_1, \dots \Lie_\phi f_n$ are all in $\Sch(\Sigma)_{\perp}$
and $e_1,\dots,e_m$ are all in $\Sch(\Sigma)_{inv}$.

Then $\Psi$ may be written
\begin{equation}
\Psi[\varphi] =
F(\varphi(\Lie_\phi f_1),\dots,\varphi(\Lie_\phi f_n),
\varphi(e_1),\dots,\varphi(e_m))
\end{equation}
for some smooth $F$.  Because $\Phi \in \Cyl_\sim$ it follows that
$F(0,\dots,0,y_1,\dots,y_m) = 0$ for all $y_1, \dots, y_n$.
For each $i\in \{1,\dots,n\}$, define
\begin{equation}
G_i(x_1,\dots,x_n,y_1,\dots,y_m)
:= \frac{F(0,\dots, 0, x_i,\dots,y_m)
- F(0,\dots, 0, x_{i+1},\dots,y_m)}
{x_i}
\end{equation}
Since $F$ is smooth, it follows that all the $G_i$ are smooth.
The $G_i$'s thus determine elements of $\Cyl$:
\begin{equation}
\Phi_i[\varphi]:=G_i(\varphi(\Lie_\phi f_1),\dots,\varphi(\Lie_\phi f_n),
\varphi(e_1),\dots,\varphi(e_m))
\end{equation}
One can also show
\begin{equation}
F \equiv \sum_{i=1}^n x_i G_i
\end{equation}
It therefore follows that
\begin{equation}
\Psi[\varphi] = \sum_{i=1}^n \varphi(\Lie_\phi f_i) \Phi_i
\end{equation}
proving the desired form.
\finishproof}

It then easily follows that
\begin{theorem}
\begin{eqnarray}
\nonumber
\Cyl^*_A &=& \{\eta \in \Cyl^* \mid \eta(\Psi)=0 \quad
\forall \Psi\in\Cyl_\sim \}
\\
\text{i.e.} \qquad \qquad
\Cyl^*_A &=& \{\eta \in \Cyl^* \mid \mathrm{Supp} \, \eta \subseteq
\scrS'(\Sigma)_{inv} \}
\end{eqnarray}
\end{theorem}
{\startproof

($\subseteq$)

Suppose $\hat{\varphi}(\Lie_\phi f)^* \eta = 0$ for all $f$.
Then $\eta(\varphi(\Lie_\phi f)\Phi)=0$ for all $f\in \Sch(\Sigma)$
and $\Phi\in\Cyl$, whence
\begin{equation}
\eta\left(\sum_{i=1}^n \varphi(\Lie_\phi f_i)\Phi_i\right)=0
\end{equation}
for all $\{f_i\} \subset \Sch(\Sigma)$ and $\{\Phi_i\}
\subset \Cyl$. The above lemma then implies $\eta$ is zero
on $\Cyl_\sim$.

($\supseteq$)

Suppose $\eta \in \Cyl^*$ is zero on $\Cyl_\sim$.  Then by the lemma,
in particular $\eta(\varphi(\Lie_\phi f)\Phi) = 0$
for all $f\in\Sch(\Sigma)$ and $\Phi \in \Cyl$, whence $\eta \in \Cyl^*_A$.
\finishproof}

\noindent
Thus, in a precise sense, $\Cyl^*_A$ is the subspace of $\Cyl^*$
consisting in elements with support only on symmetric configurations.

Next, let us construct an embedding of the Hilbert space of
the reduced theory, $\Hil_{red}$, in $\Cyl^*_A$.

For $\Psi \in \Hil_{red}$, define $\mathfrak{E}(\Psi) \in \Cyl^*$
by
\begin{equation}
\label{alpha_def}
\mathfrak{E}(\Psi)[\Phi]:= \langle \Psi, \Phi \circ I^{-1} \rangle.
\end{equation}
$\mathfrak{E}: \Hil_{red} \rightarrow \Cyl^*$ is then manifestly
anti-linear.

Note that for
$\Phi = \Phi_{red} \otimes \Phi_{\perp} \in \Cyl$,
with $\Phi_{red} \in \Hil_{red}$ and $\Phi_{\perp} \in \Hil_{\perp}$,
$\Phi \circ I^{-1} = \Phi_{\perp}[0]\Phi_{red}$,
so that in this case, (\ref{alpha_def}) can be rewritten
\begin{equation}
\label{alpha_eq1}
\mathfrak{E}(\Psi)[\Phi]=\langle \Psi, \Phi_{red}\rangle \Phi_{\perp}[0]
\end{equation}
Thus, using the standard embedding of $\Hil_{red}$ into $\Cyl_{red}^*$
using the inner product, the action of $\mathfrak{E}$ may equivalently be written
\begin{equation}
\label{alpha_eq2}
\mathfrak{E}(\Psi) = \Psi^* \otimes \delta,
\end{equation}
where $\delta: \Phi_{\perp} \mapsto \Phi_{\perp}[0]$ is the
Dirac measure, and the meaning of the notation on the right
hand side is clear from (\ref{alpha_eq1}).  It will be convenient in
this section to let $\Cyl_{\perp}$ denote the space of
cylindrical functions on $\Sch'(\Sigma)_{\perp}$; then
$\delta \in \Cyl_{\perp}^*$.

Note that (\ref{alpha_eq2}) makes it manifest that
$\mathfrak{E}$ is one to one and is an embedding of $\Hil_{red}$ in
$\Cyl^*$. Furthermore,
\begin{theorem}
The image of $\mathfrak{E}$ is contained in $\Cyl^*_A$.
\end{theorem}
{\startproof

For all $\Psi \in \Hil_{red}$, $f \in \scrS(\Sigma)$, and $\Phi \in \Cyl$,
\begin{equation}
\mathfrak{E}(\Psi)\left( \hat{\varphi}[\Lie_\phi f] \Phi \right)
= \langle \Psi, (\hat{\varphi}[\Lie_\phi f] \Phi) \circ I^{-1} \rangle
\end{equation}
Now, for $\varphi \in \scrS'(\Sigma)_{inv}$,
\begin{equation}
(\hat{\varphi}[\Lie_\phi f] \Phi)[\varphi]
=\varphi(\Lie_\phi f)\Phi[\varphi]=0
\end{equation}
whence $(\hat{\varphi}[\Lie_\phi f]\Phi)\circ I^{-1} = 0 $ for all
$\Phi \in \Cyl$.
So
\begin{equation}
\mathfrak{E}(\Psi)\left(\hat{\varphi}[\Lie_\phi f] \Phi \right)
= 0 \quad \forall \Phi \in \Cyl, f\in \scrS(\Sigma)
\end{equation}
whence
\begin{equation}
\hat{\varphi}[\Lie_\phi f]^*\mathfrak{E}(\Psi) = 0 \quad \forall f.
\end{equation}
so that $\mathfrak{E}(\Psi) \in \Cyl^*_A$ for all $\Psi \in \Hil_{red}$.
\finishproof}
Thus $\mathfrak{E}$ gives an anti-linear embedding
of $\Hil_{red}$ in $\Cyl^*_A$.  Let the image of this
embedding be denoted $\Hil_A$.



\begin{theorem}
\label{res1}
$\Hil_A \subsetneq \Cyl^*_{inv}$.
\footnote{
Ideally one would have liked to prove the stronger result
$\Cyl^*_A \subsetneq \Cyl^*_{inv}$: but in fact one does
not even have $\Cyl^*_A \subseteq \Cyl^*_{inv}$.  One has to
restrict to $\Hil_A$ before one has a subspace of $\Cyl^*_{inv}$.
This reminds us of the importance of restricting to appropriately
defined normalizable states before expecting certain
properties to hold.
}
\end{theorem}
{\startproof

First we prove $\Hil_A \subseteq \Cyl^*_{inv}$.

For all $\Psi \in \Hil_{red}$, $g\in \symgr$, and $\Phi \in \Cyl$,
\begin{eqnarray}
\nonumber
(g \cdot \mathfrak{E}(\Psi))(\Phi)&:=& \mathfrak{E}(\Psi)(g^{-1}\cdot \Phi)
\\
&=& \langle \Psi, (g^{-1} \cdot \Phi)\circ I^{-1}\rangle.
\end{eqnarray}
Now, for all $f \in \scrS'(B)$,
\begin{eqnarray}
\nonumber
((g^{-1}\cdot \Phi)\circ I^{-1})[f]&:=& (g^{-1}\cdot \Phi)[I^{-1}(f)]:=\Phi[g \cdot (I^{-1}(f))]
\\
&=& \Phi[I^{-1}(f)] = (\Phi\circ I^{-1})[f]
\end{eqnarray}
whence $(g^{-1}\cdot \Phi)\circ I^{-1} = \Phi \circ I^{-1}$, and
\begin{equation}
(g \cdot \mathfrak{E}(\Psi))(\Phi) = \langle \Psi,\Phi \circ I^{-1} \rangle = \mathfrak{E}(\Psi)(\Phi)
\end{equation}
whence $\mathfrak{E}(\Psi) \subseteq \Cyl^*_{inv}$ for all $\Psi \in \Hil_{red}$.

Next, to show $\Hil_A \subsetneq \Cyl^*_{inv}$,
we construct an element of $\Cyl^*_{inv}$ that is
not in $\Hil_A$.

To facilitate explicit calculation, let us choose
\begin{equation}
f(\rho,z,\phi):= H(\rho,z) \sin \phi
\end{equation}
where $H(\rho,z)$ is any non-negative, non-zero, smooth
function of compact support such that all derivatives of
$H$ vanish at $\rho=0$ (to ensure smoothness of $f$ at the axis).

Define $\alpha \in \scrS'(\Sigma)$ by
\begin{equation}
\alpha(h):= \int_{\Sigma} (h f)\dif^3 x
\end{equation}
Then define $\eta \in \Cyl^*$ by
\begin{equation}
\eta(\Phi):= \int_{g\in \symgr} \Phi[g\cdot \alpha]\dif g .
\end{equation}
so that $\eta \in \Cyl^*_{inv}$.

To show $\eta \notin \Cyl^*_A$, we construct an element
$\Phi$ of $\Cyl_\sim$ such that $\eta(\Phi)\not=0$.

Let $F(x):=x^2$, so that $F$ is smooth, zero only
at zero, and positive everywhere else.  Define $\Phi \in \Cyl$
by
\begin{equation}
\Phi[\varphi]:=F(\varphi(\Lie_\phi f))
\end{equation}
so that $\Phi$ is in $\Cyl_\sim$.

We have
\begin{equation}
\eta(\Phi) = \frac{1}{2\pi} \int_{\phi' = 0}^{2\pi} F(\alpha(g(-\phi')\cdot \Lie_\phi f))\dif \phi'
\end{equation}
where we have parametrized the group of rotations $\symgr$ in the usual
way by $\phi' \in \frac{\R}{2\pi\Z}$.
Working out the expression further, we get
\begin{equation}
\eta(\Phi) =
\frac{1}{2\pi} \int_{\phi' = 0}^{2\pi} F(-a\pi \sin \phi')\dif \phi'
\end{equation}
where $a:= \int_B H(\rho,z)^2 \rho \dif \rho \dif z > 0$.  Since the above integrand is positive
almost everywhere, $\eta(\Phi)>0$.

Thus $\eta \notin \Cyl^*_A$, proving in particular
$\eta \notin \Hil_A$, so that $\Hil_A \subsetneq \Cyl^*_{inv}$.
\finishproof}

Lastly, we make some remarks as to the (dual) action
of the Hamiltonian $\hat{\mathbb{H}}$
and the lack of preservation of the A-symmetric sector by
$\hat{\mathbb{H}}^*$.
As mentioned earlier, $\hat{\mathbb{H}}$ in (\ref{schr_ham})
preserves $\Cyl$; thus it has a has a dual action $\hat{\mathbb{H}}^*$
on $\Cyl^*$. We will show that, as expected from the classical analysis
(see section \ref{sect_classanal})
, $\hat{\mathbb{H}}^*$
does not preserve $\Hil_A$.  In fact, this lack of preservation is maximal:
$\hat{\mathbb{H}}^*$ maps every (non-zero) element of $\Hil_A$ out of
$\Hil_A$.  We proceed to prove this, starting with a lemma.
\begin{lemma}
$\delta \in \Cyl_{\perp}^*$ is not an eigenstate of
$\hat{\mathbb{H}}_\perp^*$.
\end{lemma}
{\startproof

Let $\lambda \in \C$ be given. We will show
$\hat{\mathbb{H}}^* \delta \neq \lambda \delta$.

Let $e$ be any non-zero element of $\Sch(\Sigma)_{\perp}$.
Let $a$ be any complex number not equal to $\frac{-\lambda}{(e,e)}$.
Define $\Phi \in \Cyl_{\perp}$ by
\begin{equation}
\Phi[\varphi]:= 1 + a(\varphi(e))^2
\end{equation}
Then, performing an explicit calculation,
\begin{equation}
(\hat{\mathbb{H}}_\perp^* \delta)(\Phi)
= \delta(\hat{\mathbb{H}}_\perp \Phi)
= - a (e,e)
\end{equation}
but
\begin{equation}
\lambda \delta(\Phi) = \lambda
\end{equation}
so that
$(\hat{\mathbb{H}}_\perp^* \delta)(\Phi) \neq \lambda \delta(\Phi)$,
proving $\hat{\mathbb{H}}_\perp^* \delta \neq \lambda \delta$ for all
$\lambda \in \C$.
\finishproof}

\begin{theorem}
$\hat{\mathbb{H}}^*$ maps every (non-zero) element of $\Hil_A$ out
of $\Hil_A$.
\end{theorem}
{\startproof

We use the fact that
$\Hil_A = \Hil_{red} \otimes \delta$ (\ref{alpha_eq2})
and we use theorem (\ref{separable_ham}).
For $\Psi \otimes \delta \in \Hil_A$,
\begin{eqnarray}
\nonumber
\hat{\mathbb{H}}^* (\Psi \otimes \delta)
&=& (\hat{\mathbb{H}}_{red}^* \otimes \ident
+ \ident \otimes \hat{\mathbb{H}}_{\perp}^*)(\Psi \otimes \delta)
\\
&=& (\hat{\mathbb{H}}_{red}^* \Psi) \otimes \delta
+ \Psi \otimes (\hat{\mathbb{H}}_{\perp}^* \delta).
\end{eqnarray}
However, as proven in the above lemma,
$\hat{\mathbb{H}}_{\perp}^* \delta$ is not again proportional to
$\delta$, whence
$\hat{\mathbb{H}}^* (\Psi \otimes \delta)$ is not
again in $\Hil_A$.
\finishproof}

In essence, then, the reason $\hat{\mathbb{H}}^*$ fails to preserve
$\Hil_A$ is that $\delta$ is not an eigenstate of
$\hat{\mathbb{H}}_\perp^*$.

It is additionally worthwhile to note that
$\hat{\mathbb{H}}^*$ also fails to preserve the
larger space $\Cyl^*_A$.  This can be seen as follows.
Let $\eta \in \Cyl^*$ be of the form $\eta(\Phi):= \Phi[\alpha]$
for some non-zero $\alpha \in \Sch'(\Sigma)_{inv}$.  One can show
from (\ref{cyla_def}) and (\ref{schr_ham}) that
$\eta \in \Cyl^*_A$ but $\hat{\mathbb{H}}^* \eta \notin \Cyl^*_A$.
Thus $\hat{\mathbb{H}}^*$ fails to
preserve $\Cyl^*_A$ in addition to failing to preserve $\Hil_A$.

\section{Analysis of $\Hil_B$}
\label{hil_s_anal}

We next analyze the structure and properties of $\Hil_B$,
helping us to (further) grasp its physical meaning in different ways.

\subsection{Reformulations of $\Hil_B$ inspired by Fock space structure}

We begin with a first reformulation of $\Hil_B$ shedding
additional light on its meaning.  Let $\hil_{inv}$ denote
the $\symgr$-invariant subspace of the single particle
Hilbert space $\hil$, and let $\hil_{inv}^\perp$ denote its
orthogonal complement.

\begin{theorem}
Suppose $\psi^{A_1 \dots A_n} \in \ntensor{n}_s \hil$.
Then
\begin{equation}
\label{hyp_1}
\overline{\xi}_{A_1} \psi^{A_1 \dots A_n} = 0
\quad \forall \xi \in \hil_{inv}^\perp
\end{equation}
iff
\begin{equation}
\psi^{A_1 \dots A_n} \in \ntensor{n}_s \hil_{inv}
\end{equation}
\end{theorem}
{\startproof

$(\Leftarrow)$: obvious.

$(\Rightarrow)$:

Let $\{v_i\}_{i\in I}$ be an orthonormal basis of $\hil_{inv}$ and
$\{v_i\}_{i\in J}$ an orthonormal basis of $\hil_{inv}^\perp$, so that
$\{v_i\}_{i\in I\cup J}$ is a basis of $\hil$.
Decomposing $\psi^{A_1 \dots A_n}$ with respect to
this basis,
\begin{equation}
\label{decomp_psi}
\psi^{A_1 \dots A_n} =: \sum_{i_1 \dots i_n \in I \cup J}
\psi^{i_1 \dots i_n} v_{i_1}^{A_1}\cdots v_{i_n}^{A_n}.
\end{equation}
Suppose (\ref{hyp_1}) holds, so that
\begin{equation}
\label{expand_hyp}
\sum_{i_1,\dots,i_n \in I \cup J}
\psi^{i_1 \dots i_n} \overline{\xi}_{A_1} v_{i_1}^{A_1}\cdots v_{i_n}^{A_n}
= 0 \quad \forall \xi \in \hil_{inv}^\perp.
\end{equation}
Suppose $\{j_1,\dots,j_n\} \not\subseteq I$.
Without loss of generality assume $j_1 \not\in I$.
Applying (\ref{expand_hyp}) in the case $\xi = v_{j_1}$ gives
\begin{equation}
\sum_{i_2,\dots,i_n \in I \cup J}
\psi^{j_1, i_2 \dots i_n} v_{i_2}^{A_2}\cdots v_{i_n}^{A_n}
= 0
\end{equation}
The linear independence of the products of basis vectors
in this expression implies
$\psi^{j_1 \dots j_n} = 0$

Thus $\psi^{j_1 \dots j_n} = 0$ for all $\{j_1, \dots j_n\} \not\subseteq I$,
whence, from (\ref{decomp_psi}),
$\psi^{A_1 \dots A_n} \in \ntensor{n}_s \hil_{inv}$.
\finishproof}

\noindent
It then follows trivially from the definition of $\Hil_B$,
the creation operators, annihilation operators and $\Psi_0$ that
\begin{corollary}
\label{symm_modes_fock}
$\Hil_B = \mspan \{ a^\dagger(\xi_1) \cdots a^\dagger(\xi_n) \Psi_0
\mid \xi_1, \dots \xi_n \in \hil_{inv} \}$ where $\Psi_0$ is the Fock vacuum.
\end{corollary}
\noindent
(Here, as throughout this paper, the span is understood to
mean the Cauchy completion of finite linear combinations of elements
of a given set.)
This gives us our first reformulation of $\Hil_B$;
it tells us that $\Hil_B$ is the space
of states in which \textit{all non-symmetric modes are unexcited}.
(In the language of \cite{bt} the non-symmetric modes
are ``quantum mechanically suppressed.'')

Next recall that, in free Klein-Gordon theory,
with each $\xi\in \hil$ one has
an associated (normalized) \textit{coherent state}
$\Psi^{coh}_\xi \in \Hil$ defined by
\begin{equation}
\label{coh_def}
\Psi^{coh}_\xi = \rme^{\hat{\Lambda}(\xi)} \Psi_0
\end{equation}
where $\hat{\Lambda}(\xi):= a^\dagger(\xi)-a(\xi)$.

In the case where $\xi \in \Gamma \subset \hil$,
$\Psi^{coh}_\xi$ has the interpretation of being the
quantum state that ``best approximates'' the classical
state $\xi$.  The expectation values of field operators
determined by $\Psi^{coh}_\xi$ are precisely the values of
the fields in $\xi$, and uncertainties in appropriate
field components are minimized.

\begin{theorem}
\label{cohstate_res}
$\Hil_B = \mspan\{ \Psi^{coh}_\xi \mid \xi \in \hil_{inv} \}$
\end{theorem}
{\startproof

Let $\Hil^{coh}_s := \mspan\{ \Psi^{coh}_\xi \mid \xi \in \hil_{inv} \}$.

For all $\xi \in \hil_{inv}$,
\begin{eqnarray}
\nonumber
\rme^{\hat{\Lambda}(\xi)}\Psi_0
&=& \rme^{(a^\dagger(\xi)-a(\xi))}\Psi_0
\\
\nonumber
&=& \rme^{a^\dagger(\xi)}\rme^{-a(\xi)}
\rme^{-\sshalf \langle \xi, \xi \rangle}\Psi_0
\\
&=& \rme^{-\sshalf \langle \xi, \xi \rangle}
\sum_{n=0}^\infty \frac{1}{n!}(a^\dagger(\xi))^n\Psi_0
\end{eqnarray}
which is in $\Hil_B$ by corollary \ref{symm_modes_fock}.
Thus $\Hil^{coh}_s \subseteq \Hil_B$

Going in the other direction, for all $\xi_1, \dots, \xi_n \in \hil_{inv}$,
\begin{eqnarray}
a^\dagger(\xi_1)\cdots a^\dagger(\xi_n)\Psi_0
&=&
\left.\tderiv{}{\lambda_1}\right|_{\lambda_1=0} \cdots
\left.\tderiv{}{\lambda_n}\right|_{\lambda_n=0}
\rme^{a^\dagger(\lambda_1\xi_1+\dots+\lambda_n\xi_n)}\Psi_0
\\
\nonumber
&=&
\left.\tderiv{}{\lambda_1}\right|_{\lambda_1=0} \cdots
\left.\tderiv{}{\lambda_n}\right|_{\lambda_n=0}
\rme^{\sshalf \parallel \lambda_1 \xi_1 + \dots + \lambda_n\xi_n\parallel^2}
\rme^{\hat{\Lambda}(\lambda_1\xi_1 + \dots + \lambda_n\xi_n)}\Psi_0
\end{eqnarray}
which is a limit of linear combinations of symmetric coherent states,
whence it is in $\Hil^{coh}_s$, so that $\Hil_B \subseteq \Hil^{coh}_s$.
\finishproof}

Note that because $\Gamma_{inv}$ is dense in $\hil_{inv}$ and
$\xi \mapsto \Psi^{coh}_\xi$ is continuous
\footnote{
The continuity of $\xi \mapsto \Psi^{coh}_\xi$ can be seen from the
relation
\begin{equation}
\nonumber
\parallel \Psi^{coh}_\xi - \Psi^{coh}_{\xi_i} \parallel^2
=
2 - 2 \cos(\mathrm{Im}\langle \xi, \xi_i \rangle)
\rme^{-\half \parallel \xi - \xi_i \parallel^2}
\end{equation}
If $\xi_i \rightarrow \xi$, from the continuity in $\xi_i$ of the right
hand side of the above equation,
$\Psi^{coh}_{\xi_i} \rightarrow \Psi^{coh}_\xi$.
}
one can replace $\hil_{inv}$ with $\Gamma_{inv}$ in the statement
of the above theorem.  The theorem then expresses $\Hil_B$ as a span
of coherent states associated with the axisymmetric sector
of the strictly classical theory.

As a side note, it is not hard to show that $\Psi^{coh}_\xi$
satisfies the usual property of being
a simultaneous eigenstate of the annihilation operators:
\begin{equation}
a(\eta) \Psi^{coh}_\xi = \langle \eta, \xi \rangle \Psi^{coh}_\xi
\end{equation}
One can use this property to prove theorem
(\ref{cohstate_res}) in an alternative way.

\subsection{Reformulations of $\Hil_B$ inspired by the structure
$\Hil = \Hil_{red} \otimes \Hil_{\perp}$;
natural isomorphism between $\Hil_{red}$ and $\Hil_B$}
\label{sect_tensref}

\begin{theorem}
\label{symm_modes_schr}
In terms of the structure $\Hil = \Hil_{red} \otimes \Hil_{\perp}$,
\begin{equation}
\Hil_B = \{ \Upsilon \otimes 1 \mid \Upsilon \in \Hil_{red} \}.
\end{equation}
\end{theorem}
{\startproof

We first show that, in terms of
$\Hil = \Hil_{red} \otimes \Hil_{\perp}$, for $[f,g] \in \hil_{inv}$,
$a^\dagger([f,g]) = \sqrt{2\pi}
a^\dagger_{red}([f,\rho g]) \otimes \ident$,
where $a^\dagger_{red}(\cdot)$ is the creation operator in the
reduced theory and $\rho$ is the spatial coordinate that is
distance from the z-axis.

From (\ref{schr_cr_op}), for $[f,g] \in \hil_{inv}$ and
$\Psi \in \Hil$, we have
\begin{eqnarray}
\nonumber
a^\dagger([f,g])\Psi[\varphi] &=&
\int_\Sigma \dif^3 x \left\{
\left(\Theta^\half f + \rmi g \right) \varphi -
\half \left(f + \rmi \Theta^{-\half}g\right)
\frac{\delta}{\delta \varphi}
\right\}\Psi[\varphi]
\\
\nonumber
&=&
\int_\Sigma \dif^3 x \left\{
\left(\Theta^\half f + \rmi g \right) \varphi_s -
\half \left(f + \rmi\Theta^{-\half}g\right)
\frac{\delta}{\delta \varphi_s}
\right\}\Psi[\varphi]
\\
=
&(2\pi)&
\int_B \dif^2 x \rho \left\{
\left(\Theta^\half f + \rmi g \right) \varphi_s -
\half \left(f + \rmi \Theta^{-\half}g\right)
\frac{\delta}{\delta \varphi_s}
\right\}\Psi[\varphi]
\end{eqnarray}
Using equations (\ref{redfield_defs}),(\ref{fderiv_relation})
and (\ref{redcr_op}),
\begin{eqnarray}
\nonumber
a^\dagger([f,g])\Psi[\varphi]
&=&
\sqrt{2\pi}\left\{\int_B \dif^2 x
\left(\rho \Theta^\half f + \rmi \rho g\right) \varphi_r
-\half \int_B \dif^2 x
\left(f + \rmi \Theta^{-\half}g\right)
\frac{\delta}{\delta \varphi_r}
\right\}\Psi[\varphi]
\\
\nonumber
&=&
\sqrt{2\pi}
\left\{\varphi_r\left[\rho \Theta^\half f + \rmi\rho g\right]
-\half \int_B \dif^2 x
\left(f + \rmi\Theta^{-\half}g\right)
\frac{\delta}{\delta \varphi_r}
\right\}\Psi[\varphi]
\\
&=&
\left(\sqrt{2\pi}a_{red}^\dagger([f,\rho g]) \otimes
\ident\right) \Psi[\varphi]
\end{eqnarray}
where the $\rho$ and $2\pi$ factors arise from our
conventions in (\ref{redfield_defs}).

Using this,
\begin{eqnarray}
\nonumber
\Hil_B &=&
\mspan \left\{ a^\dagger([f_1,g_1]) \cdots a^\dagger([f_n,g_n]) (1\otimes 1)
\right\}_{\substack{
f_1,\dots f_n,\hfill \\
g_1,\dots g_n \in \scrS(\Sigma)_{inv}}}
\\
\nonumber
&=&
\mspan \left\{
\left( (2\pi)^{\frac{n}{2}}a_{red}^\dagger([f_1,\rho g_1]) \cdots
a_{red}^\dagger([f_n,\rho g_n]) 1 \right)\otimes 1
\right\}_{\substack{
f_1,\dots f_n,\hfill \\
g_1,\dots g_n \in \scrS(B)}}
\\
\nonumber
&=&
\mspan \left\{ a_{red}^\dagger([f_1,\rho g_1]) \cdots
a_{red}^\dagger([f_n,\rho g_n]) 1
\right\}_{\substack{
f_1,\dots f_n,\hfill \\
g_1,\dots g_n \in \scrS(B)}}\otimes 1
\\
&=& \Hil_{red} \otimes 1
\end{eqnarray}

\finishproof}

\noindent
From this it is obvious that $\Hil_B$ and $\Hil_{red}$ are
naturally isomorphic.

Note that, in contrast to the $\delta$ in (\ref{alpha_eq2}),
$1$ \textit{is} an eigenfunction of $\hat{\mathbb{H}}_{\perp}$:
$1$ is the unique vacuum of $\hat{\mathbb{H}}_{\perp}$.
A number of important conclusions follow from this fact.
\begin{enumerate}
\item
First, Using (\ref{sep_ham_eq}), it is now easy to see
that $\hat{\mathbb{H}}$ preserves $\Hil_B$.
Because $\hat{\mathbb{H}}$ preserves $\Hil_B$, it induces, via
the isomorphism between $\Hil_B$ and $\Hil_{red}$, an operator on $\Hil_{red}$.
It is easy to see that this induced Hamiltonian on $\Hil_{red}$ is just
$\hat{\mathbb{H}}_{red}$, so that everything is consistent.

\item
Because
$1$ is the vacuum of $\hat{\mathbb{H}}_{\perp}$, the above theorem
gives another expression of the fact that $\Hil_B$ is the
space of states in which ``all non-symmetric modes are unexcited''.

\item
Because $1$ is the unique eigenstate of
$\ident \otimes \hat{\mathbb{H}}_{\perp}$ with eigenvalue zero,
the above theorem implies
\begin{equation}
\Hil_B = \Ker \left(\ident \otimes \hat{\mathbb{H}}_{\perp}\right)
\end{equation}
Thus, $\ident \otimes \hat{\mathbb{H}}_{\perp}$ by itself could have been
taken as the sole constraint.
\end{enumerate}
The significance of the last point is the following. One can cast
$\hat{\mathbb{H}}_{\perp}$ as a quadratic combination of the
\textit{original second class set of self-adjoint constraint
operators} (equation (\ref{secclass_constr})).  This in turn makes
this way of imposing the (ideal) constraints (\ref{secclass_constr})
an instance of Klauder's universal procedure for imposing
constraints (with `$\delta$' being set to zero; see \cite{klauder}).

Let us show this. First let $\{f_i\}$ denote
any basis of $\scrS(\Sigma)_{\perp}$ orthonormal with respect to
$(\cdot, \Theta^\half \cdot)$.  Then
$\{\xi_i:= (f_i,\Theta^\half f_i)\}$ is an orthonormal basis of $h_{\perp}$, the orthogonal complement
of the axisymmetric subspace in the single particle Hilbert space $h$
(as one can check).
Define
\begin{eqnarray}
\hat{\eta}_{[i,0]} &:=& \hat{\varphi}[f_i]
\\
\hat{\eta}_{[i,1]} &:=& \hat{\pi}[f_i].
\end{eqnarray}
so that $\{\hat{\eta}_{[i,A]}\}$ is a basis of the full original
set of constraint operators (\ref{secclass_constr}).
Define the matrix
\begin{equation}
M^{[i,A],[j,B]}:= \alpha^{AB} \langle \xi_i, \hat{H} \xi_j \rangle
= \alpha^{AB} (f_i, \Theta f_j)
\end{equation}
where
$\alpha^{AB}
:= \scriptsize \half
\left(
\begin{array}{cc}
1  & i \\
-i & 1
\end{array} \right)$.
It is not hard to check that
$M$ is Hermitian and positive definite.
We have
\begin{equation}
\ident \otimes \hat{\mathbb{H}}_{\perp}
= \sum_{[i,A],[j,B]} M^{[i,A][j,B]}
\hat{\eta}_{[i,A]} \hat{\eta}_{[j,B]}
\end{equation}
casting $\ident \otimes \hat{\mathbb{H}}_{\perp}$ in the desired form.

Additionally from theorem \ref{symm_modes_schr}, it is also easy to see
\begin{corollary}
\label{res2}
$\Hil_B \subsetneq \Hil_{inv}$.
\end{corollary}
{\startproof

That $\Hil_B \subseteq \Hil_{inv}$ is immediate
from theorem (\ref{symm_modes_schr}) and the fact that
the symmetry group $\symgr$ acts non-trivially only on the
second factor in $\Hil = \Hil_{red}\otimes \Hil_{\perp}$.

To show that furthermore $\Hil_B \subsetneq \Hil_{inv}$
note that there exist non-constant $\symgr$-symmetric
elements of $\Hil_\perp$.  If $\alpha$ is one of these
elements of $\Hil_\perp$, and $\psi\in\Hil_{red}$ arbitrary
and non-zero,$\psi \otimes \alpha$ is in $\Hil_{inv}$ but not
$\Hil_B$.
\finishproof}

\noindent
In addition to the example used in the proof above,
one can also give a more ``concrete'' example of an element of
$\Hil_{inv}$ that is not in $\Hil_B$: a two particle state in which the
two particles are in z-angular momentum eigenstates with equal and
opposite eigenvalue (i.e. ``spin up'' and ``spin down'' eigenstates).
It is easy to see that such a state is in $\Hil_{inv}$:
its total z-angular momentum is zero and so it is annihilated by
$\hat{\mathbb{L}}_z$.  It is also not too hard to show it is
not in $\Hil_B$.

\subsection{Minimization of fluctuations from axisymmetry}
\label{fluctsubsect}

A last notable property of $\Hil_B$ is that the fluctuations from axisymmetry
in its members are under complete control and are in a certain sense
\textit{minimized}, whereas in $\Hil_{inv}$ there is no control over
fluctuations from axisymmetry.

Let us be more precise.  Recall ideally one may wish to impose
$\Lie_\phi \hat{\varphi}(x) \Psi =0$ and
$\Lie_\phi \hat{\pi}(x)\Psi = 0$,
but that, in this form, this is not possible.
Therefore we imposed instead a complex linear
combination of these constraints (in approach B).
Nevertheless, the resulting
states $\Psi \in \Hil_B$ are still such that
\begin{eqnarray}
\label{symexp_conf}
\langle \Psi, \Lie_\phi \hat{\varphi}(x) \Psi \rangle &=& 0
\\
\label{symexp_mom}
\langle \Psi, \Lie_\phi \hat{\pi}(x) \Psi \rangle &=& 0
\end{eqnarray}
that is, the expectation values of $\hat{\varphi}(x)$ and
$\hat{\pi}(x)$ are
axisymmetric.
The easiest way to see this is actually to first note
that $\Hil_B \subset \Hil_{inv}$ and then show that
(\ref{symexp_conf}) and (\ref{symexp_mom}) hold for
\textit{all} members of $\Hil_{inv}$. One can show this
using the fact that for all
rotations $g$,
$\hat{\varphi}(g\cdot x) = U_g \hat{\varphi}(x) U_g^{-1}$.

Thus both $\Hil_B$ and $\Hil_{inv}$ consist in states giving rise to
axisymmetric field expectation values. The difference between
$\Hil_B$ and $\Hil_{inv}$ comes, however, when we consider
\textit{fluctuations} from axisymmetry.

To show this, it will be convenient to first note that if
$\varphi_s(x), \varphi_{\perp}(x)$ denote the symmetric and
non-symmetric parts of $\varphi(x)$, and
$\pi_s(x), \pi_{\perp}(x)$ denote the symmetric and non-symmetric
parts of $\pi(x)$, so that $\hat{\varphi}_s(x), \hat{\pi}_s(x)$
are operators on $\Hil_{red}$ and
$\hat{\varphi}_{\perp}(x), \hat{\pi}_{\perp}(x)$ are operators
on $\Hil_{\perp}$, we have
\begin{eqnarray}
\hat{\varphi}(x) &=& \hat{\varphi}_s(x) \otimes \ident
+ \ident \otimes \hat{\varphi}_{\perp}(x)
\\
\hat{\pi}(x) &=& \hat{\pi}_s(x) \otimes \ident
+ \ident \otimes \hat{\pi}_{\perp}(x)
\end{eqnarray}
For any operator $\hat{O}$ on $\Hil$ and $\Psi \in \Hil$, the ``fluctuation''
in $\hat{O}$ determined by $\Psi$ is defined by
\begin{equation}
\Delta_\Psi \hat{O} := \sqrt{\langle \Psi, \hat{O}^2 \Psi \rangle - \langle \Psi, \hat{O} \Psi \rangle^2}
\end{equation}
Smearing the symmetry constraint operators against a test function $f$, we get
$\hat{\varphi}[\Lie_\phi f]$ and $\hat{\pi}[\Lie_\phi f]$.
For $\Psi = \Upsilon \otimes 1 \in \Hil_B$, with unit norm, one can show
the uncertainties in the non-axisymmetric modes are given by
\begin{eqnarray}
\label{phispread}
\Delta_\Psi \hat{\varphi}[\Lie_\phi f]
&=& \sqrt{\half \int \dif^3 x (\Lie_\phi f) \Theta^{-\half} \Lie_\phi f}
\\
\label{pispread}
\Delta_\Psi \hat{\pi}[\Lie_\phi f]
&=& \sqrt{\half \int \dif^3 x (\Lie_\phi f) \Theta^\half \Lie_\phi f}
\end{eqnarray}
In particular, for $f$ an eigenfunction of $\Theta$,
\begin{equation}
\Delta_\Psi \hat{\varphi}[\Lie_\phi f] \Delta_\Psi \hat{\pi}[\Lie_\phi f] = \half
\end{equation}
saturating Heisenberg's uncertainty principle.
\footnote{
There exists a complete basis of eigenfunctions $f$ of $\Theta$.
However, technically $\hat{\varphi}[f]$ and $\hat{\pi}[g]$ are well defined as operators
only when $f$ and $g$ are in $\scrS(\Sigma)$ --- and no eigenstates of $\Theta$ are in
$\scrS(\Sigma)$.  Therefore, \textit{prima facie} the spreads
$\Delta_\Psi \hat{\varphi}[\Lie_\phi f]$ and $\Delta_\Psi \hat{\pi}[\Lie_\phi f]$
are not defined for eigenfunctions $f$.
Nevertheless, the right hand sides of equations
(\ref{phispread}) and (\ref{pispread}) \textit{are} well defined
for $f$ an eigenfunction of $\Theta$, so that we can take the
spreads to be simply defined by these expressions in that case.
}

\section{Carrying operators from $\Hil$ to $\Hil_{red}$}
\label{sect_induced_ops}

We have finished investigating the properties of A and B symmetry
in the present simple model.

One of the nice properties of $\Hil_B$ is that
the Hamiltonian preserves it, so that the Hamiltonian has a well-
defined restriction to $\Hil_B$ which can then be carried over
to $\Hil_{red}$ via the natural isomorphism.  The operator
thereby induced on $\Hil_{red}$ is the same as the Hamiltonian
in the reduced theory, so that $\Hil_B$ gives a fully dynamical
embedding of the reduced theory.

However, in more general situations, even if the Hamiltonian preserves
a given choice of ``symmetric sector'' in a given theory,
other operators of interest may not.  It is therefore of
interest to investigate the possibility of a general rule for
carrying over \textit{any} operator $\hat{\mathcal{O}}$ on
the full theory Hilbert space $\Hil$ to an operator
$\hat{\mathcal{O}}_{red}$ on the reduced theory Hilbert space
$\Hil_{red}$ that somehow ``best approximates the information
contained in $\hat{\mathcal{O}}$.''  We will motivate and
suggest such a prescription for a completely general theory, and
then look at applications to example operators in the model theory
considered in this paper.  We
assume only that we are given some embedding $\iota$ of the reduced theory,
$\Hil_{red}$ into the full theory $\Hil$.
\footnote{
This is yet another advantage of B symmetry, at least in the present
simple model: the states are normalizable, and it is only in this
case that the general prescription described here will apply.
In the case of A symmetry, even though the Hamiltonian does not
preserve $\Hil_A$, one might have still hoped to
induce a Hamiltonian operator on $\Hil_{red}$ from that on
$\Hil$ via some other manner, such as the one described here; 
but it is not at all obvious
how to do that due to the non-normalizability of A-symmetric
states.  The combination of the Hamiltonian not preserving $\Hil_A$
and $\Hil_A$ not having any normalizable elements thus
frustrates attempts to use A symmetry to compare
the dynamics in the full and reduced theories in any systematic way.}

For $\hat{\mathcal{O}}$ Hermitian (\textit{i.e.}, symmetric), a list of
physically desireable criteria for the corresponding $\hat{\mathcal{O}}_{red}$
might include
\begin{enumerate}
\item
\label{hermicpres}
$\hat{\mathcal{O}}_{red}$ is Hermitian.
\item
\label{matrixpres}
$\langle \iota \Psi_1, \hat{\mathcal{O}} \iota \Psi_2 \rangle = \langle \Psi_1, \hat{\mathcal{O}}_{red} \Psi_2 \rangle$
\item
\label{fluctpres}
$\Delta_{\iota\Psi_1} \hat{\mathcal{O}} = \Delta_{\Psi_1} \hat{\mathcal{O}}_{red}$
\end{enumerate}
for all $\Psi_1, \Psi_2$ in $\Hil_{red}$.  That is, one might want
Hermicity, matrix elements and fluctuations to be preserved.

Fortunately the second of these criteria uniquely determines $\hat{\mathcal{O}}_{red}$:
\begin{equation}
\label{redop_def}
\hat{\mathcal{O}}_{red}:=
\iota^{-1} \circ \scrP \circ \hat{\mathcal{O}} \circ \iota.
\end{equation}
where $\scrP: \Hil \rightarrow \iota[\Hil_{red}]$
denotes orthogonal projection.
This is perhaps what one would first write down as a possible
prescription.  The point, however, is that this prescription is not 
\textit{ad hoc}:
it is uniquely determined by a physical criterion.  Furthermore,
\begin{theorem}
The prescription defined in (\ref{redop_def}) satisfies \textit{all three}
of the desired properties, except that the last property is replaced by
\begin{equation}
\label{fluct_rel}
\Delta_{\iota\Psi_1} \hat{\mathcal{O}} \ge \Delta_{\Psi_1} \hat{\mathcal{O}}_{red}
\end{equation}
with equality holding iff $\hat{\mathcal{O}}$ preserves $\iota[\Hil_{red}]$.
\footnote{
As a side note, this result fully extends to non-Hermitian operators
if we replace condition (\ref{hermicpres}) with
$(\hat{\mathcal{O}}_{red})^\dagger = (\hat{\mathcal{O}}^\dagger)_{red}$,
define the spread of a non-Hermitian operator by
\begin{equation}
\Delta_\Psi \hat{\mathcal{O}}
:= \sqrt{\langle \Psi,
\thalf ( \hat{\mathcal{O}}^\dagger \hat{\mathcal{O}}
+ \hat{\mathcal{O}} \hat{\mathcal{O}}^\dagger) \Psi \rangle
- \mid \langle \Psi, \hat{\mathcal{O}} \Psi \rangle \mid^2 }.
\end{equation}
and give as the condition for equality in (\ref{fluct_rel})
the condition that \textit{both}
$\hat{\mathcal{O}}$ and $\hat{\mathcal{O}}^\dagger$ preserve
$\iota[\Hil_{red}]$.
}
\footnote{
Even though Hermicity of $\hat{\mathcal{O}}$ implies
$\hat{\mathcal{O}}_{red}$ is Hermitian, \textit{self-adjointness}
of $\hat{\mathcal{O}}$ does not imply self-adjointness
of $\hat{\mathcal{O}}_{red}$.  This fact is discussed on
page 19 of \cite{bs}.}
\end{theorem}

Let us look at some example operators in the Klein-Gordon theory
considered in the present paper.  The example of
$\hat{\mathbb{H}}$ has already been remarked upon.
We proceed, then, to look at the basic configuration and momentum
operators $\hat{\varphi}[f], \hat{\pi}[g]$.
It is convenient to split these operators into parts.
Define as operators on $\Hil$,
\begin{eqnarray}
\hat{\tilde{\varphi}}_s[f] := \hat{\varphi}[f_s] &\qquad&
\hat{\tilde{\pi}}_s[g] := \hat{\pi}[g_s]
\\
\hat{\tilde{\varphi}}_{\perp}[f] := \hat{\varphi}[f_\perp] &\qquad&
\hat{\tilde{\pi}}_\perp[g] := \hat{\pi}[g_\perp]
\end{eqnarray}
Note that the latter pair of operators are just the symmetry constraint
operators. The tildes on these four operators are to
distinguish them from
the related operators on $\Hil_{red}$ and $\Hil_\perp$.
For the configuration operators we have
$\hat{\tilde{\varphi}}_s[f] \Psi [\varphi] = \varphi_s[f] \Psi[\varphi]$
and $\hat{\tilde{\varphi}}_{\perp}[f] \Psi[\varphi] = \varphi_\perp[f] \Psi[\varphi]$.
In terms of the corresponding operators on
$\Hil_{red}$ and $\Hil_\perp$,
\begin{eqnarray}
\hat{\tilde{\varphi}}_s[f] := \hat{\varphi}_s[f] \otimes \ident  &\qquad&
\hat{\tilde{\pi}}_s[g] := \hat{\pi}_s[g]\otimes \ident
\\
\hat{\tilde{\varphi}}_{\perp}[f] := \ident \otimes \hat{\varphi}_\perp[f]
&\qquad&
\hat{\tilde{\pi}}_\perp[g] := \ident \otimes \hat{\pi}_\perp[g]
\end{eqnarray}
$\hat{\tilde{\varphi}}_s[f]$ and $\hat{\tilde{\pi}}_s[g]$
both preserve $\Hil_B$, whereas $\hat{\tilde{\varphi}}_\perp[f]$
and $\hat{\tilde{\pi}}_\perp[g]$ do not. Nevertheless, on carrying these
operators over to the reduced theory using (\ref{redop_def}) we get exactly
what one would expect:
\begin{eqnarray}
(\hat{\tilde{\varphi}}_s[f])_{red} &=& \hat{\varphi}_s[f]
\\
(\hat{\tilde{\pi}}_s[g])_{red} &=& \hat{\pi}_s[g]
\end{eqnarray}
but
\begin{eqnarray}
(\hat{\tilde{\varphi}}_\perp[f])_{red} &=& 0
\\
(\hat{\tilde{\pi}}_\perp[g])_{red} &=& 0 .
\end{eqnarray}
The inclusion of orthogonal projection in prescription
(\ref{redop_def}) is essential in getting the last couple
of equations above.  Even though the symmetry constraint
operators $\hat{\tilde{\varphi}}_\perp[f] = \hat{\varphi}[f_\perp]$
and $\hat{\tilde{\pi}}_\perp[g] = \hat{\pi}[g_\perp]$ do not
annihilate $\Hil_B$, nevertheless, as one would hope, their
corresponding operators induced on $\Hil_{red}$ via (\ref{redop_def})
are identically zero.

Lastly, one can also look at the angular momentum operator
$\hat{\mathbb{L}}_z$.  In the reduced classical theory
the $z$-angular momentum is identically zero, so that one
would expect the corresponding operator to be identically zero
as well.  Indeed,
\begin{equation}
(\hat{\mathbb{L}}_z)_{red}
:= \iota^{-1} \circ \scrP \circ \hat{\mathbb{L}}_z \circ \iota
= 0
\end{equation}
as follows from $\Hil_B \subset \Hil_{inv}$.

\section{Summary and outlook}

\subsection{Physical meaning(s) of $\Hil_B$}

It is notable that $\Hil_B$ has a number of characterizations
with completely distinct physical meaning all pointing to
ways in which $\Hil_B$ embodies the notion of ``symmetry.''
They are

\begin{enumerate}
\item
\label{constr_char} $\Hil_B$ is the solution space to a set of
constraints whose classical analogues isolate the axisymmetric
sector of the classical phase space;
\item
$\Hil_B$ is the span of the coherent states associated with the axisymmetric
sector of the classical theory;
\item
$\Hil_B$ is the space of states in which all non-symmetric modes are unexcited.
In terms of the Fock picture, this characterization of $\Hil_B$ took the form
of corollary \ref{symm_modes_fock}
and in terms of the Schr\"{o}dinger picture this characterization took the
form of theorem \ref{symm_modes_schr}.
\end{enumerate}

The first two of these in a clear way point to $\Hil_B$
as the ``quantum analogue of the \textit{classical} axisymmetric
sector''. The idea of invariance under the group action (leading to
$\Hil_{inv}$), on the other hand, is the quantum analogue of classical
axisymmetry in a slightly more indirect sense.  It is \textit{invariance}
under the quantum analogue of \textit{classical rotation about the z axis}.
It is a subtle but clear distinction.  Another way to state this distinction is
that in $\Hil_{inv}$ we are imposing `$\Lie_{\phi} \Psi = 0$',
whereas in $\Hil_B$
we are imposing (an appropriate complex linear combination of) the conditions
$\Lie_{\phi} \varphi(x)=0$, $\Lie_{\phi} \pi(x)=0$.
In $\Hil_{inv}$
we are imposing axisymmetry on the \textit{wave-function} whereas in $\Hil_B$
we are imposing axisymmetry on the \textit{field operators}.

One can see the distinction in yet another way as well.
Recall in the classical theory that the total angular momentum
is given by the expression
\begin{equation}
\mathbb{L}_z = \int_{\Sigma} \pi(\Lie_\phi \varphi) \dif^3 x
\end{equation}
Classically the condition $\mathbb{L}_z = 0$ is weaker than
the condition that $\Lie_\phi \varphi = 0$ and $\Lie_\phi \pi = 0$.
Likewise, as theorem \ref{res1} and corollary \ref{res2} showed us,
quantum mechanically $\hat{\mathbb{L}}_z \Psi = 0$ is weaker
than (an appropriate reformulation of)
$\Lie_\phi \hat{\varphi}(x) \Psi = 0$ and
$\Lie_\phi \hat{\pi}(x) \Psi = 0$.  Again, it is $\Hil_B$ (and $\Hil_A$)
that is playing the role of the quantum analogue(s) of classical axisymmetry.

Furthermore,
as was seen in section \ref{fluctsubsect},
one can grasp the difference between
$\Hil_{inv}$ and $\Hil_B$ in terms of fluctuations from axisymmetry.
Expectation values for field operators are axisymmetric both for states
in $\Hil_{inv}$ and for states in $\Hil_B$.  However, the standard deviation, or ``fluctuations'',
of $\hat{\varphi}[\Lie_\phi f]$ and $\hat{\pi}[\Lie_\phi g]$ from zero
are completely controlled in $\Hil_B$, whereas in $\Hil_{inv}$ one has no
control over these fluctuations.

Lastly, it is $\Hil_B$ and $\Hil_A$ that achieve commutation
of symmetry reduction and quantization, the former at the full level of
dynamics.  $\Hil_{inv}$ does not achieve commutation at any level.

\subsection{Future directions: sketch of application to LQG}
\label{LQGapp}

As pointed out earlier, the embedding of symmetry reduced
theories into full loop quantum gravity suggested by Bojowald is analogous
to the embedding $\Hil_A$ in the Klein-Gordon model considered here.
Nevertheless, in the Klein-Gordon model, we saw that,
for multiple reasons, the embedding $\Hil_B$
is preferable to $\Hil_A \subseteq \Cyl^*_A$:
\begin{enumerate}
\item
$\hat{\mathbb{H}}$ preserves $\Hil_B$ whereas $\hat{\mathbb{H}}^*$ does not
preserve $\Hil_A$.  Consequently, it is only $\Hil_B$
that gives us an embedding of \textit{both} Hilbert space structure
and dynamics.

\item
Fluctuations from axisymmetry in $\Hil_B$ are more evenly distributed
between configuration and momentum variables, and are in a certain
sense minimized.
\item
$\Hil_B$ is the span of the set of coherent states associated with the
symmetric sector of the classical theory --- a particularly elegant
characterization that brings out a physical content not shared by $\Hil_A$.
\end{enumerate}

It would be ideal, then, if one could
extend the notion of B-symmetry (embodied in $\Hil_B$)
to the case of LQG.
The most obvious avenue for this is to use the characterization
in theorem \ref{cohstate_res} --- that of the span of
semi-classical states associated
with the symmetric sector of the classical theory.  For, ideas on
semi-classical states in LQG have already been introduced
\cite{compl_coh, al, conrady}.
Indeed, one of the results in \cite{bt} seems to partially
support this strategy.  There it was found that one
had to restrict precisely to coherent symmetric states before one could
reproduce in full LQG a result known in the reduced theory --- namely,
the boundedness of the inverse volume operator.
\footnote{
The following is a side note.
\cite{bt} nevertheless found that, on more general states
approximately invariant under the action of the symmetry group
(translations and rotations) on large scales,
the inverse volume operator is \textit{unbounded}.
From this  they conclude that ``the boundedness
of the inverse scale factor
in isotropic and homogeneous LQC does not extend to the full
theory even when restricting LQG to those states which one
would use to describe a maximally homogeneous and isotropic
situation (modulo fluctuations)''(pp.4-5).\\
\indent
However, in light of the present research, as written, this statement
is not wholly just.
For, as was pointed out earlier, in quantum gravity, the notion of
symmetry given simply by invariance under the action of the symmetry
group becomes trivial once one goes to the level of solutions to the
diffeomorphism constraint.  Therefore, the symmetry restriction used in
\cite{bt} to make the statement of unboundedness is, strictly speaking,
empty of physical content.
Rather, as has been a main point of this paper, when comparing
a full quantum theory with a corresponding symmetry reduced theory,
the notion of symmetric sector in the full theory should be
more restrictive than the one defined by invariance under the
symmetry group action. And, on the choice of ``symmetric sector''
suggested by the present research, \cite{bt} did find boundedness.
}

However, there is a freedom in the choice of semiclassical states
used in defining $\Hil_B^{LQG}$. Let us consider how this
freedom can be used to reproduce additional characteristics of
B-symmetry.

First, characterization (\ref{constr_char}) listed in the last
subsection is easily
reproduced by using complexifier coherent states.
To see this, let $P$ denote the $\SU(2)$ principal bundle for
the theory, with base space $\Sigma$. Let $\mathcal{S}$,
a subgroup of the automorphisms of $P$,
be the symmetry group of interest.
Then if we define $\Hil^{LQG}_B$ to be the span of complexifier
coherent states associated with symmetric field configurations,
all states $\Psi$ in $\Hil^{LQG}_B$ will satisfy
\begin{equation}
\label{lqg_symm_constr}
\widehat{\rule{0em}{1.2em}
\left(\Phi_{\alpha}^* A^\C(e) - A^\C(e)\right)}\Psi =
\left(U_{\alpha}\hat{A}^\C(e)U_{\alpha}^{-1} - \hat{A}^\C(e)\right)\Psi
=0
\end{equation}
for all edges $e$ and all
$\alpha \in \mathcal{S} \subset \Aut(P)$.
Here $\Phi_{\alpha}$ and $U_{\alpha}$ denote the action of $\alpha$
on the kinematical phase space and kinematical Hilbert space, respectively.
$\hat{A}^\C(\cdot)$ are the ``annihilation operators'' defined in
\cite{compl_coh} depending on a particular choice of
complexifier.
The classical constraints under the hat on the left hand side
select uniquely, at the classical level, the
\mbox{($\scrS$-)symmetric}
sector.\footnote{
assuming the $A^\C(e)$ for all edges $e$ separate points in the
kinematical phase space.  This is gauranteed to be true at least
locally on the phase space and is hoped to hold globally
for complexifiers of physical interest \cite{compl_coh}.
}
So, again like $\Hil_B$
in the scalar field case, $\Hil^{LQG}_B$ will solve
a set of constraints that, at the classical level,
uniquely select the appropriate classical symmetric sector.

Perhaps more importantly, one would like to reproduce the property
that $\Hil_B$ is preserved by the Hamiltonian (in the case of LQG, a
constraint in the bulk). It is not obvious how to do this;
nevertheless we mention some possibilities. Perhaps complexifier
coherent states could again be used, with the complexifier being
`tailored' to the dynamics in some way; or perhaps one needs a
different approach. Essentially
what one needs is \textit{`temporally stable'} or
\textit{`dynamical'} coherent states if $\Hil_B$ is to be preserved
by the Hamiltonian constraint.  This can be seen as follows. Let
$\Gamma$ denote the classical phase space for a given theory, and
let $\Hil$ denote the corresponding quantum state space. Define a
family of coherent states $F: \Gamma \rightarrow \Hil$ (associating
each classical phase space point with a quantum state) to be
temporally stable if there exists a map $A: \Gamma \times \mathbb{R}
\rightarrow \Gamma$ such that
\begin{equation}
e^{-it\hat{H}} F(p) = F(A(p,t))
\end{equation}
for all $p \in \Gamma$ and $t \in \mathbb{R}$.
Suppose we are given a group $G$ with action on both $\Gamma$ and
$\Hil$.  Suppose the Hamiltonian $\hat{H}$ is $G$-invariant, and the
family of coherent states $F$ is chosen to be $G$-covariant.  It
is not hard to see that $A(p,t)$ will be $G$-covariant as well.
If we then define $\Hil_B$ to be the span
of all $F(p)$ for $p \in \Hil$ fixed by $G$,
$\hat{H}$ will preserve $\Hil_B$, as desired.

The problem of constructing temporally stable coherent states is discussed in
\cite{klauder_coh1, klauder_coh2, wb}.  In \cite{klauder_coh1} and
\cite{klauder_coh2}, two general schemes are given for constructing
stable families of coherent states. Unfortunately in both of these schemes,
the label space for the coherent states is no longer necessarily the
classical phase space, $\Gamma$, whence it is not obvious whether it
is possible or appropriate to use such coherent states in
constructing $\Hil_B^{LQG}$ in the manner described above.
In \cite{wb}, on the other hand, $\Gamma$ is retained as the label
space for the coherent states, but they conclude that
\textit{exactly} stable families of coherent states do not always exist, but
rather, for interacting theories, one in general expects only
\textit{approximately} stable families. However, this
statement is made for a fixed set of `fundamental operators' used to
characterize semi-classicality; it is not clear it holds if the
choice of `fundamental operators' is not so fixed.

We leave investigation along these lines to future research. The
main reason for desiring $\Hil_B^{LQG}$ to be preserved by the
Hamiltonian constraint is that then a constraint operator
$\hat{C}(x)_{red}$ is induced on $\Hil_B^{LQG}$, making
$(\Hil_B^{LQG}, \hat{C}(x)_{red})$ a closed system that could be
compared, for example, with loop quantum cosmology. However, even if
$\Hil_B^{LQG}$ is not preserved by dynamics, $\Hil_B^{LQG}$ is still
valuable in that it gives us a notion of `symmetric sector'.
This notion of `symmetric sector' can in principle be transferred 
to the physical
Hilbert space (as described below), at which point preservation by
constraints is no longer an issue.  Comparison with LQC might then
be attempted directly at the level of the physical Hilbert space
\cite{lqc_phys}.

Next let us discuss two issues related to constraints. First, as
just touched upon, is the question of how one might obtain from
$\Hil^{LQG}_B$ a `symmetric sector' in the final \textit{physical}
Hilbert space of LQG. Let $\Hil_{kin}$ denote the kinematical
Hilbert space of the theory, let $\Hil_{Diff}$ denote the solution
to the Gauss and diffeomorphism constraints, and let $\Hil_{Phys}$
denote the space solving the Hamiltonian constraint as well.  We
have already suggested how to define the ``B-symmetric sector'' in
$\Hil_{kin}$. To obtain a notion of symmetric sector in
$\Hil_{Diff}$, the obvious strategy is to group average the
``B-symmetric states'' in $\Hil_{kin}$. This strategy is natural in
light of \cite{abc} and the fact that we are using the definition of
the symmetric sector inspired by theorem \ref{cohstate_res}.
Furthermore, if one follows the master constraint programme
\cite{mastconstr}, one can use the master constraint to group
average\footnote{
or, more or less equivalently, use the zero eigenvalue spectral
projection operator for the master constraint
}
states from $\Hil_{Diff}$ to $\Hil_{Phys}$ and so transfer
the notion of symmetric sector to $\Hil_{Phys}$.

The second issue related to constraints is that of gauge-fixing
the symmetry group --- that is, choosing a symmetry group which is
not invariant under conjugation by diffeomorphisms and gauge
transformations.  Such a choice of symmetry group is made in
LQC, for example.  We note the following: on group averaging
the symmetric sector over gauge transformations and diffeomorphisms,
any such gauge-fixing will be washed out. This can be seen as follows.
Let $\Hil_B^G$ denote the ``symmetric sector''
of $\Hil_{kin}$ corresponding to the subgroup $G$ of the automorphism
group of the principal bundle.  If the only ``background'' used
in the construction of $\Hil_B^G$ is the choice of group $G$, then,
for any automorphism $\alpha$ of $P$, we will have
covariance:
\begin{equation}
U_\alpha [\Hil_B^G] = \Hil_B^{\alpha \cdot G \cdot \alpha^{-1}}.
\end{equation}
Now, if we had not gauge fixed, our symmetric sector would
consist in the span of all $\Hil_B^{\alpha \cdot G \cdot \alpha^{-1}}$ for
$\alpha$ in the automorphism group.
This follows from the fact that we are defining the quantum
symmetric sector as the span of coherent states associated with the
classical symmetric sector.
Thus, from the above equation, it is clear that on group
averaging over the automorphism group, one will obtain the same subspace of
$\Hil_{Diff}$ whether one gauge fixes the symmetry group or doesn't.

Indeed, this situation can be
mimicked in the Klein-Gordon toy model by simply declaring,
for example, that $\mathbb{L}_x, \mathbb{L}_y, \mathbb{L}_z$
be constraints.  This is a first class system, and the gauge group
generated is the full group of $\SO(3)$ rotations about the origin.
In this context,
the group of rotations about the z-axis is then
\begin{enumerate}
\item a subgroup of the full canonical gauge group.
\item furthermore a gauge-fixed group.  It is not left invariant
by conjugation by the rest of the canonical gauge group.
\end{enumerate}
These two properties precisely mimic the situation in loop
quantum cosmology.
In this toy model, one has the possibility check that certain nice
properties of $\Hil_B$ are preserved by the group averaging procedure,
such as the minimization of fluctuations from axisymmetry.
This could possibly be done by group averaging the kinematical
symmetry constraints
$\{ \hat{\varphi}[\Lie_\phi f], \hat{\pi}[\Lie_\phi g]\}$
to obtain operators on the physical Hilbert space.  One could
then calculate the fluctuations of these operators from zero
for the proposed symmetric sector in the physical Hilbert space.

\textit{A final note.} What has mainly been discussed thus far is how
one should define the notion of ``symmetric sector'' in LQG
appropriate for comparison with reduced models. It is not at all
clear, however, whether one should expect the ``symmetric sector''
so defined to be isomorphic to the Hilbert space in the corresponding
model quantized a la Bojowald.
\footnote{
There is a fundamental difference between the configuration
algebra underlying the full theory and the configuration algebra
underlying LQC and other Bojowald-type models.
Specifically: in the Bojowald-type models, only holonomies
along edges adapted to the symmetry are
included in the algebra.
This makes isomorphism with $\Hil_B^{LQG}$ seem less likely
unless perhaps $\Hil_B^{LQG}$ is modified in some way.
}.
If it is not, we argue that the physics of the ``symmetric sector''
defined along the lines suggested in this section
should be considered the ``more fundmantal'' description.
Perhaps one could even formulate the physics of this
sector in such a way that one could easily calculate
corrections to predictions made using Bojowald-type models
such as LQC.

\section*{Acknowledgements}

The author would like to thank Abhay Ashtekar for suggesting the
initial direction for this research and for helpful discussions,
Martin Bojowald and Florian Conrady for helpful discussions,
and Daniele Perini for raising an initial
question in part leading to this research.
In addition, the author thanks Abhay Ashtekar and Florian Conrady
for looking over a draft of this paper, and thanks the referees for
their comments which resulted in significant improvements in the
final draft.
This work was supported in part
by NSF grant PHY-0090091, the Eberly Research Funds of Penn State,
and the Frymoyer and Duncan Fellowships of Penn State.

\appendix

\section{The reduced-then-quantized free scalar field theory}
\label{redtheory}

Let $(\rho,z,\phi)$ denote standard cylindrical coordinates on $\Sigma$
such that the symmetry vector field $\phi^a$ is equal to $\pderiv{}{\phi}$.
Let \label{redman2}$B:=\Sigma / \symgr$ denote the reduced
spatial manifold.
Let $P:\Sigma \rightarrow B$ denote canonical projection, and let
$q^{ab}:= P_* g^{ab}$. B may be coordinatized by $(\rho, z)$, which are then
Cartesian coordinates for $q_{ab}:=(q^{ab})^{-1}$.
The configuration and momentum variables $\varphi$ and $\pi$ may then be
represented by functions on B.

More specifically, for $\varphi$ and $\pi$ symmetric, we define
\footnote{
The definition of $\pi_r$ can be motivated by considering
the weight one densitization of $\pi$, $\tilde{\pi}:=(\det g)^{\half} \pi$.
Using the projection mapping $P: \Sigma \rightarrow B$,
we can then define $\tilde{\pi}_r := \sqrt{2\pi} P_* \tilde{\pi}$,
where the push-forward is defined by treating $\tilde{\pi}$ as a measure.
If we then dedensitize $\tilde{\pi}_r$ using $q_{ab}$:
$\pi_r:= (\det q)^{-\half} \tilde{\pi}_r$, then
\begin{equation}
\pi_r = \sqrt{2\pi} \rho \pi.
\end{equation}
}
\begin{equation}
\label{redfield_defs}
\varphi_r(\rho,z) = \sqrt{2\pi}\varphi(\rho,z) \qquad
\pi_r(\rho,z) = \sqrt{2\pi}\rho\pi(\rho,z)
\end{equation}
(The $\sqrt{2\pi}$ factors are included for later convenience.)
Let $\Gamma_{red}$ denote the reduced phase space -- the space
of all possible $[\varphi_r,\pi_r]$
\footnote{
Classically there are also boundary conditions which
$\varphi_r$ and $\pi_r$ must satisfy at $\rho=0$ in order
to ensure smoothness.  However, when going over to the quantum
theory, because there is no surface term in the symplectic
structure at $\rho=0$, there are no separate degrees of
freedom at $\rho=0$, and the boundary conditions do not matter.
}.
The symplectic structure induced on $\Gamma_{red}$ is simply
\begin{equation}
\label{redsylstr}
\Omega([\varphi_r,\pi_r], [\varphi_r',\pi_r'])
= \int_B (\pi_r \varphi_r' - \varphi_r \pi_r') \dif \rho \dif z
\end{equation}

Thus we see that at least kinematically, in terms of $(\varphi_r, \pi_r)$,
the reduced theory is nothing other than a free Klein-Gordon
theory on B with flat metric $q_{ab}$.
\footnote{
Note the role of the definition of $\varphi_r$ and $\pi_r$ in
making this the case.
}
From the time evolution of
$(\varphi, \pi)$ in the full theory, the time-evolution of $(\varphi_r, \pi_r)$
is
\begin{eqnarray}
\label{redevol_phi}
\dot{\varphi}_r &=& \rho^{-1} \pi_r
\\
\nonumber
\dot{\pi}_r &=& \rho (\Delta_{\Sigma} - m^2) \varphi_r
\\
\label{redevol_pi}
&=& \rho (\Delta_B + \frac{1}{\rho} \pderiv{}{\rho} - m^2) \varphi_r
\end{eqnarray}
where $\Delta_{\Sigma}$ denotes the Laplacian on $\Sigma$ determined by
$g_{ab}$ and $\Delta_B$ denotes the Laplacian on $B$ determined by $q_{ab}$. Let
$\Theta:= -\Delta_{\Sigma}+m^2 = -\Delta_B -\frac{1}{\rho}\pderiv{}{\rho} + m^2$.
Note that from $(\Theta^q f,g)_\Sigma = (f, \Theta^q g)_\Sigma$ for arbitrary
$q \in \Q$, it follows $(\rho \Theta^q f,g)_B = (f, \rho \Theta^q g)_B$.

Given a choice of parametrization of time, from \cite{qft_refs1},
the naturally associated complex structure on the classical phase space is
\begin{equation}
J = -(-\Lie_{\xi}\Lie_{\xi})^{-\half} \Lie_{\xi}
\end{equation}
where $\Lie_{\xi}$ denotes derivative with respect to the time evolution
vector field $\xi$.  From (\ref{redevol_phi}, \ref{redevol_pi}), one then
calculates
\begin{equation}
J [\varphi_r, \pi_r] = [-\Theta^{-\half}\rho^{-1}\pi_r, \rho \Theta^\half \varphi_r ].
\end{equation}
Following \cite{qft_refs1}, the Hermitian inner product thereby determined on the
classical phase space is
\begin{equation}
\langle [\varphi_r,\pi_r], [\varphi_r',\pi_r'] \rangle
= \half(\rho \Theta^\half \varphi_r, \varphi'_r)_B
+ \half(\Theta^{-\half}\rho^{-1}\pi_r,\pi'_r)_B
-\frac{\rmi}{2}(\pi_r,\varphi'_r)_B+\frac{\rmi}{2}(\varphi_r,\pi'_r)_B
\end{equation}
where $(f,g)_B:= \int_B fg \dif \rho \dif z$
\footnote{
Unless otherwise specified, from now on all integrations
over $B$ are understood to be with respect to
$\dif^2 x := \dif \rho \dif z$,
and all integrations over $\Sigma$ are understood to be with
respect to $\dif^3 x := \rho \dif \rho \dif z \dif \phi$.
}
.  We take the quantum configuration space to be
$\scrS'(B)$, with quantum measure given, again following \cite{qft_refs1}, by
\begin{equation}
\text{``} \dif \mu_{red} =
\exp\left\{-\half(\varphi, \rho \Theta^\half \varphi)_B\right\}
\scrD \varphi. \text{''}
\end{equation}
More rigorously, the Fourier transform of the measure is given by
\begin{equation}
\chi_{\mu_{red}}(f) = \exp\left\{-\half(f,\Theta^{-\half} \rho^{-1} f)_B\right\}
\end{equation}
We will denote the space of cylindrical functions in the reduced theory
by $\Cyl_{red}$. That is, $\Cyl_{red}$ is the space of functions
$\Phi:\scrS'(B) \rightarrow \C$ of the form
\begin{equation}
\Phi[\alpha] = F(\alpha(f_1),\dots,\alpha(f_n))
\end{equation}
for some $f_1, \dots, f_n \in \scrS(B)$ and some
smooth $F: \R^n \rightarrow \C$ with growth less than
exponential.

The representation of the field observables
$\varphi[f]:= \int_B f\varphi$ and $\pi[g]:= \int_B g \pi$ is given by
\begin{eqnarray}
\label{redphi_op}
(\hat{\varphi}_r[f] \Psi)[\varphi_r] &=& \varphi_r[f] \Psi[\varphi_r]
\\
\label{redpi_op}
(\hat{\pi}_r[g] \Psi)[\varphi_r] &=& -\rmi\int_B
\left(g\frac{\delta}{\delta\varphi_r} - \varphi_r \rho \Theta^\half g\right)
\Psi[\varphi].
\footnotemark
\end{eqnarray}
\footnotetext{
As noted earlier, $\frac{\delta}{\delta \varphi_r}$ is defined with respect
to the volume form $\dif \rho \dif z$.
}
For a given point $[\varphi_r,\pi_r]=[f,g]$ in the classical
phase space, we have the ``classical observables'' for the corresponding
annihilation and creation operators:
\begin{eqnarray}
\nonumber
a_{red}([f,g])\mid_{[\varphi_r,\pi_r]} &=& \langle [f,g],[\varphi_r,\pi_r] \rangle
\\
&=& \half(\varphi_r[\rho\Theta^\half f-\rmi g]
+\pi_r[\Theta^{-\half}\rho^{-1}g+\rmi f])
\\
\nonumber
a^\dagger_{red}([f,g])\mid_{[\varphi_r,\pi_r]}
&=& \langle [\varphi_r,\pi_r],[f,g] \rangle
\\
&=& \half(\varphi_r[\rho\Theta^\half f+\rmi g]+
\pi_r[\Theta^{-\half}\rho^{-1}g-\rmi f])
\end{eqnarray}
Quantizing by substituting in (\ref{redphi_op}, \ref{redpi_op}), we obtain
\begin{eqnarray}
\label{redcr_op}
a_{red}^\dagger([f,g]) &=&
\varphi_r[\rho \Theta^\half f + \rmi g]
-\frac{\rmi}{2}\int_B\left\{\Theta^{-\half}(\rho^{-1}g)- \rmi f\right\}
\frac{\delta}{\delta \varphi_r}
\\
\label{redan_op}
a_{red}([f,g]) &=&
-\frac{\rmi}{2}\int_B\left\{\Theta^{-\half}(\rho^{-1}g)+ \rmi f\right\}
\frac{\delta}{\delta \varphi_r}
\end{eqnarray}
Lastly we quantize the (reduced) Hamiltonian.  The reduced Hamiltonian is
\begin{equation}
\label{redham}
\mathbb{H}_{red} = \half \int_B
( \rho^{-1}\pi_r^2 + \rho(\vec{\nabla}\varphi_r)^2 + \rho m^2 \varphi_r^2 )
\dif \rho \dif z
\end{equation}
This can be checked to be consistent with (\ref{redsylstr}, \ref{redevol_phi},
\ref{redevol_pi}).
We next rewrite the Hamiltonian,
\begin{eqnarray}
\mathbb{H}_{red} &=& \half \int_B (\rho^{-1}\pi_r^2 + \rho (\vec{\nabla}\varphi_r)^2
+ \rho m^2 \varphi_r^2)\dif \rho \dif z
\\
&=& \half \int_B \left( \rho^{-1}\pi_r^2
+ \rho \varphi_r \left\{ -\Delta_B - \rho^{-1} \pderiv{}{\rho} + m^2 \right\}\varphi_r \right)
\dif \rho \dif z
\\
&=& \half \int_B (\rho^{-1} \pi_r^2 + \rho \varphi_r \Theta \varphi_r)\dif \rho \dif z
\end{eqnarray}
From (\ref{redevol_phi},\ref{redevol_pi}), we deduce the single particle Hamiltonian:
\begin{eqnarray}
\hat{H}_{red} [\varphi_r,\pi_r] &=& J \deriv{}{t} [\varphi_r,\pi_r]
\\
&=& [ \Theta^\half \varphi_r, \rho \Theta^\half \rho^{-1} \pi_r ]
\end{eqnarray}
So,
\begin{equation}
\mathbb{H}_{red} = \langle [\phi_r,\pi_r], \hat{H}_{red} [\phi_r,\pi_r] \rangle
\end{equation}
matching one's expectations.
Let $\{\xi_i = [f_i,g_i]\}$ denote an arbitrary basis of $\Gamma_{red}$,
orthonormal with respect to $\langle \cdot, \cdot \rangle$. Then,
\begin{eqnarray}
\mathbb{H}_{red} &=& \sum_{i,j} \langle [\phi_r,\pi_r], \xi_i \rangle
\langle \xi_i, \hat{H}_{red} \xi_j \rangle \langle \xi_j, [\phi_r,\pi_r] \rangle
\\
&=& \sum_{i,j} \langle \xi_i, \hat{H}_{red} \xi_j \rangle
a^\dagger_{red}(\xi_i) a_{red}(\xi_j)
\end{eqnarray}
To quantize we use the normal ordering above and substitute in
(\ref{redcr_op}, \ref{redan_op}), to obtain
\begin{equation}
\mathbb{H}_{red} = \int_{\substack{x\in B},{y\in B}}
\left\{ A(x,y)
\varphi_r(y) \frac{\delta}{\delta \varphi_r(x)}
- B(x,y)
\frac{\delta^2}{\delta\varphi_r(x)\delta\varphi_r(y)} \right\}
\end{equation}
where
\begin{eqnarray}
A(x,y)
&:=&
\half \sum_{i,j} \langle \xi_i, \hat{H}_{red} \xi_j \rangle
(f_j - \rmi \Theta^{-\half} \rho^{-1} g_j)(x)
(\rho \Theta^\half f_i + \rmi g_i)(y)
\\
B(x,y)
&:=&
\frac{1}{4} \sum_{i,j} \langle \xi_i, \hat{H}_{red} \xi_j \rangle
(\Theta^{-\half} g_i - \rmi f_i)(x)
(\Theta^{-\half} \rho^{-1} g_j + \rmi f_j)(y)
\end{eqnarray}
By integrating against test functions, one can show
$A(x,y)$ is the integral kernel of $\Theta^{\half}$,
and $B(x,y) = \half \rho^{-1} \delta^2(x,y)$.  It follows
\begin{equation}
\label{reducedham}
\hat{\mathbb{H}}_{red} = \int_{x\in B} \left\{
(\Theta^\half \varphi_r)(x) \frac{\delta}{\delta \varphi_r(x)}
- \half \rho^{-1} \frac{\delta^2}{\delta \varphi_r(x)^2} \right\}.
\end{equation}

This expression is in fact equal to (\ref{ham_sympart}).
This can be seen in the following manner.
Because the fields $\varphi_r(\rho,z)$ are in one-to-one correspondence with
the axisymmetric fields $\varphi_s(\rho,z,\phi)$
($\varphi_r = (2\pi)^{-\half} \varphi_s$), functionals depending on a $\varphi_r$
can be interpreted as functionals depending on a symmetric field
$\varphi_s(\rho,z,\phi)$ and vice-versa.
Consequently $\frac{\delta}{\delta \varphi_r(\rho,z)}$
and $\frac{\delta}{\delta \varphi_s(\rho,z,\phi)}$ can be understood
to operate on the same space.  Furthermore, for all $f(\rho,z)$,
\begin{eqnarray}
\nonumber
\int_B \dif \rho \dif z f(\rho,z) \frac{\delta}{\delta \varphi_r(\rho,z)}
&=&
\frac{1}{\sqrt{2\pi}} \int_{\Sigma} \dif \rho \dif z \dif \phi \rho f(\rho,z)
\frac{\delta}{\delta \varphi_s(\rho,z,\phi)}
\\
&=&
\sqrt{2\pi} \int_B \dif \rho \dif z \rho f(\rho,z) \frac{\delta}{\delta \varphi_s(\rho,z,\phi_o)}
\end{eqnarray}
where $\phi_o$ is arbitrary.  Thus
\begin{equation}
\label{fderiv_relation}
\frac{\delta}{\delta \varphi_r(\rho,z)} =
\sqrt{2\pi} \rho \frac{\delta}{\delta \varphi_s(\rho,z,\phi_o)}.
\end{equation}
Substituting this into (\ref{ham_sympart}) then gives (\ref{reducedham}).

\section{List of symbols and basic relations}
\label{symb_app}

\newlength{\oldparindent}
\setlength{\oldparindent}{\parindent}
\setlength{\parindent}{0in}


\textit{For Klein-Gordon model}:
\\

\entry{$\Sigma$}{spatial hyperplane in Minkowski space}
\entry{$x^1, x^2, x^3$}{Cartesian coordinates on $\Sigma$}
\entry{$\rho, \phi, z$}{cylindrical coordinates on $\Sigma$}
\entry{$\Diff(\Sigma)$}{group of diffeomorphisms of $\Sigma$}
\entry{$\symgr \subset \Diff(\Sigma)$}
{the group of rotations about z-axis}
\entry{$\vec{\phi}:= \tpderiv{}{\phi}$}{axial symmetry field}
\entry{$\Lie_\phi$}{Lie derivative with respect to $\phi$}
\entry{$B:= \Sigma/\symgr$}{spatial manifold for the reduced theory}
$\dif^2 x = \dif \rho \dif z$ \\
$\dif^3 x = \rho \dif \rho \dif \phi \dif z$ \\
$(f,g) = (f,g)_\Sigma := \int_\Sigma f g \dif^3 x$ \\
$(f,g)_B := \int_B f g \dif^2 x$ \\
\entry{$\Gamma$}{full phase space}
\entry{$[f,g]$}{point in $\Gamma$ defined by $\varphi = f$, $\pi = g$.
(not to be confused with commutator; context makes clear which is intended)}
\entry{$\Gamma_{inv} \subset \Gamma$}
{$\symgr$-invariant subspace of $\Gamma$}
\entry{$\Gamma_A \subset \Gamma$}
{classical solution to constraint set A}
\entry{$\Gamma_B \subset \Gamma$}
{classical solution to constraint set B}
\vspace{-0.5cm}
\begin{eqnarray*}
\Gamma_A &\subset& \Gamma_{inv} \\
\Gamma_B &=& \Gamma_{inv}
\end{eqnarray*}
\entry{$\Omega(\cdot, \cdot)$}
{symplectic structure on $\Gamma$; in appendix
 \ref{redtheory}: symplectic structure in the reduced theory}
\entry{$\Delta = \Delta_{\Sigma}$}{Laplacian on $\Sigma$}
\entry{$\Delta_B$}{Laplacian on $B$}
\entry{$m$}{scalar field mass}
\entry{$\Theta:= -\Delta+m^2$}{}
\entry{$J$}{complex structure on $\Gamma$; in appendix \ref{redtheory}:
  complex structure in the reduced theory}
\entry{$\Sch(\Sigma)$, $\Sch(B)$}
{space of Schwarz functions on $\Sigma$, $B$}
\entry{$\Sch'(\Sigma)$, $\Sch'(B)$}
{space of tempered distributions on $\Sigma$, $B$}
\entry{$\Sch(\Sigma)_{inv}$, $\Sch'(\Sigma)_{inv}$}
{$\symgr$-invariant subspaces of $\Sch(\Sigma)$ and $\Sch'(\Sigma)$,
respectively}
\entry{$P: \Sigma \rightarrow B$}{canonical projection}
\entry{$I: \Sch'(\Sigma)_{inv} \rightarrow \Sch'(B)$}
{is defined by $[I(\beta)](f):=\beta(P^* f)$; I is an isomorphism}
\entry{$\pi$, $\Pi$}
{group averaging maps on $\Sch(\Sigma)$ and $\Sch'(\Sigma)$,
respectively (see \S \ref{sect_struct})}
\entry{$\Sch(\Sigma)_{\perp}$}
{the kernel of $\pi$; equivalently, the orthogonal
complement of $\Sch(\Sigma)_{inv}$ in $\Sch(\Sigma)$}
\entry{$\Sch'(\Sigma)_{\perp}$}{the kernel of $\Pi$}
\vspace{-0.5cm}
\begin{eqnarray*}
\Sch(\Sigma)_{inv} &\natiso& \Sch(B)\\
\Sch'(\Sigma)_{inv} &\natiso& \Sch'(B)\\
\Sch(\Sigma) &=& \Sch(\Sigma)_{inv} \oplus \Sch(\Sigma)_{\perp} \\
\Sch'(\Sigma) &=& \Sch'(\Sigma)_{inv} \oplus \Sch'(\Sigma)_{\perp}
\end{eqnarray*}
\entry{$\mu$}{quantum measure on $\Sch'(\Sigma)$}
\entry{$\mu_{red}$}{quantum measure on $\Sch'(B)$}
\entry{$\mu_{\perp}$}{unique measure on $\Sch'(\Sigma)_{\perp}$
 such that $\mu = \mu_{red} \times \mu_{\perp}$}
\entry{$h$}{single particle Hilbert space of full theory}
\entry{$\ntensor{n}$}{$n$-fold tensor product}
\entry{$\ntensor{n}_s$}{symmetrized $n$-fold tensor product}
\entry{
$\Hil:= L^2(\Sch'(\Sigma),\dif \mu)$
\\
\dummy \hspace{0.4cm} $= \sfock(h)$}{full field theory Hilbert space}
\entry{$\Hil_{inv}$}{$\symgr$-invariant subspace of $\Hil$}
\entry{$\Hil_{red}$ \\
\dummy \hspace{0.2cm} $:= L^2(\Sch'(B),\dif \mu_{red})$}
{reduced theory Hilbert space}
$\Hil_{\perp}:= L^2(\Sch'(\Sigma),\dif \mu_{\perp})$ \\
\entry{$\langle \cdot, \cdot \rangle$}
{inner product on $\hil$, $\Hil$, $\Hil_{red}$, or $\Hil_{\perp}$,
depending on context}
\entry{$\Cyl$, $\Cyl^*$}
{space of cylindrical functions in the full theory, and its algebraic dual}
\entry{$\Cyl_{red}$, $\Cyl_{red}^*$}
{space of cylindrical functions in the reduced theory, and its algebraic dual}
\entry{$\Cyl_{\perp}$, $\Cyl_{\perp}^*$}
{space of cylindrical functions on $\Sch'(\Sigma)_{\perp}$, and the
algebraic dual}
\begin{eqnarray*}
\Hil &=& \Hil_{red} \otimes \Hil_{\perp} \\
\Cyl &\hookrightarrow& \Hil \hookrightarrow \Cyl^* \\
\Cyl_{red} &\hookrightarrow& \Hil_{red} \hookrightarrow \Cyl_{red}^* \\
\Cyl_{\perp} &\hookrightarrow& \Hil_{\perp} \hookrightarrow \Cyl_{\perp}^*
\end{eqnarray*}
\entry{$\Cyl^*_{inv}$}
{$\symgr$-invariant subspace of $\Cyl^*$}
\entry{$\Cyl^*_A \subset \Cyl^*$}
{quantum mechanical solution to constraint set A}
\entry{$\mathfrak{E}: \Hil_{red} \hookrightarrow \Cyl^*$}
{is defined by $\mathfrak{E}(\Psi)[\Phi]:= \langle \Psi, \Phi \circ I^{-1} \rangle$}
\entry{$\Hil_A:= \mathrm{Im}\, \mathfrak{E} \subset \Cyl^*_A$}{}
\entry{$\Hil_B \subset \Hil$}
{quantum mechanical solution to constraint set B}
\entry{$\hil_{inv}$}
{$\symgr$-invariant subspace of $\hil$}
\entry{$\hil_{\perp}$}
{orthogonal complement of $\hil_{inv}$ in $\hil$}
\entry{$\hat{\varphi}[f]$, $\hat{\pi}[g]$}
{basic smeared field operators in the full theory}
\entry{$f_s$, $f_{\perp}$}
{components of a given $f \in \Sch(\Sigma)$
with respect to the decomposition
$\Sch(\Sigma)=\Sch(\Sigma)_{inv}\oplus \Sch(\Sigma)_{\perp}$}
\entry{$\varphi_s$, $\varphi_{\perp}$}
{components of $\varphi$ with respect to the decomposition
$\Sch(\Sigma) = \Sch(\Sigma)_{inv} \oplus \Sch(\Sigma)_{\perp}$
or $\Sch'(\Sigma) = \Sch'(\Sigma)_{inv} \oplus \Sch'(\Sigma)_{\perp}$,
according to the context}
\entry{$\pi_s$, $\pi_{\perp}$}
{components of $\pi$ with respect to the decomposition
$\Sch(\Sigma) = \Sch(\Sigma)_{inv}\oplus \Sch(\Sigma)_{\perp}$}
\entry{
$\varphi_r = \varphi_{red}$ \\
\dummy \hspace{0.5cm}$:= (2\pi)^{\sshalf} \varphi_s$, \\
$\pi_r = \pi_{red}$ \\
\dummy \hspace{0.5cm}$:= (2\pi)^{\sshalf} \rho \pi_s$}
{basic classical fields in the reduced theory; relation
to fields in the full theory.}
\entry{
$\hat{\varphi}_r[f] = \hat{\varphi}_{red}[g]$\\
$\hat{\pi}_r[g[ = \hat{\pi}_{red}[g]$
}{smeared field operators in the reduced theory}
\entry{
$\hat{\varphi}_s[f]$, $\hat{\pi}_s[g]$}
{the operators on $\Hil_{red}$ corresponding to the smeared functions
$\varphi_s[f]$ and $\pi_s[g]$.
$\hat{\varphi}_s[f]$ acts by multiplication and
$\hat{\pi}_s[g]$ is the self-adjoint part of
$-\rmi \int_{\Sigma} g \tfrac{\delta}{\delta \varphi_s}$.
}
\entry{
$\hat{\varphi}_{\perp}[f]$, $\hat{\pi}_{\perp}[g]$}
{the operators on $\Hil_{\perp}$ corresponding to
$\varphi_{\perp}[f]$ and $\pi_{\perp}[g]$.  $\hat{\varphi}_{\perp}[f]$
is defined by multiplication and $\hat{\pi}_{\perp}[g]$
is the self-adjoint part of
$-\rmi \int_{\Sigma} g \tfrac{\delta}{\delta \varphi_{\perp}}$
}
\begin{eqnarray*}
\hat{\varphi}_s[f] &=& (2\pi)^{-\sshalf}\hat{\varphi}_{red}[f_s] \\
\hat{\pi}_s[g] &=& (2\pi)^{-\sshalf}\hat{\pi}_{red}[\rho^{-1} g_s] \\
\hat{\varphi}[f]
&=& \hat{\varphi}_s[f] \otimes \ident
+ \ident \otimes \hat{\varphi}_{\perp}[f] \\
\hat{\pi}[g]
&=& \hat{\pi}_s[g] \otimes \ident + \ident \otimes \hat{\pi}_{\perp}[g]
\end{eqnarray*}
\entry{$a(\cdot)$, $a^\dagger(\cdot)$}
{annihilation and creation operators in the full theory, or their
classical counterparts, depending on the context}
\entry{
$a_r(\cdot) = a_{red}(\cdot)$\\
$a_r^\dagger(\cdot) = a_{red}^\dagger(\cdot)$}
{annihilation and creation operators in the reduced theory, or
their classical counterparts, depending on the context}
\entry{$\Psi_0$}{vacuum in the full theory (\S \ref{sect_prelim})}
\entry{$\hat{H}$, $\hat{H}_{red}$}
{single particle Hamiltonians
in the full and reduced theories,
respectively}
\entry{$\mathbb{H}$, $\hat{\mathbb{H}}$}
{total Hamiltonian in the full theory and its quantization}
\entry{$\mathbb{H}_{red}$, $\hat{\mathbb{H}}_{red}$}
{total Hamiltonian in the reduced theory and its quantization}
\entry{$\hat{\mathbb{H}}_{\perp}$}
{the unique operator on $\Hil_{\perp}$ such that
$\hat{\mathbb{H}} = \hat{\mathbb{H}}_{red} \otimes \ident
+ \ident \otimes \hat{\mathbb{H}}_{\perp}$}
\entry{$\mathbb{L}_x$, $\mathbb{L}_y$, $\mathbb{L}_z$}
{x, y, and z-components, respectively, of the total angular momentum
in the full classical theory}
\entry{$\hat{\mathbb{L}}_z$}
{operator on $\Hil$ corresponding to $\mathbb{L}_z$}
\entry{$U_g$}{action of a given $g \in \symgr$ on $\Hil$}
\entry{$\Delta_\Psi \hat{\mathcal{O}}$}
{variance (``fluctuation'') in $\hat{\mathcal{O}}$
for the state $\Psi$}
\entry{$\hat{\Lambda}(\xi):=a^\dagger(\xi)-a(\xi)$}{}
\entry{$\Psi_\xi^{coh}:=\rme^{\hat{\Lambda}(\xi)}\Psi_0$}
{coherent state in $\Hil$ corresponding to a given $\xi \in \hil$}
\entry{$\mspan\{\cdot\}$}
{Cauchy completion of the set of all finite linear combinations}

\textit{For LQG (\S\ref{LQGapp})}:
\\

\entry{$\Sigma$}{spatial Cauchy surface}
\entry{$P$}{$\SU(2)$ principal bundle over $\Sigma$}
\entry{$\Aut(P)$}{group of automorphisms of $P$}
\entry{$\scrS \subset \Aut(P)$}{symmetry group under consideration}
\entry{$\Hil_{kin}$}{kinematical Hilbert space}
\entry{$\Hil_{Diff}$}{solution to Gauss and Diffeomorphism constraints}
\entry{$\Hil_{Phys}$}{solution to all constraints}
\entry{$\Hil_B^{LQG} \subset \Hil_{kin}$}
 {proposal for B-symmetric sector of LQG}
\entry{$\Phi_{\alpha}$}{action of $\alpha \in \Aut(P)$
  on the kinematical phase space}
\entry{$U_{\alpha}$}{action of $\alpha \in \Aut(P)$ on $\Hil_{kin}$}
\entry{$\hat{A}^\C(\cdot)$, $A^\C(\cdot)$}
{annihilation operators and their classical counterparts, as defined in
 \cite{compl_coh}}

\setlength{\parindent}{\oldparindent}


\begin{thebibliography}{99}
 \bibitem{lqg_revs}
    Ashtekar A and Lewandowski J 2004 Background independent quantum gravity:
      A status report \textit{Class. Quant. Grav.} \textbf{21} R53\\
    Rovelli C 2004 \textit{Quantum Gravity} (Cambridge: Cambridge UP)\\
    Thiemann T 2003 Lectures on loop quantum gravity
      \textit{Lect. Notes Phys.} \textbf{631} 41-135
 \bibitem{lqc_revs}
    Bojowald M 2002 Isotropic loop quantum cosmology
    \textit{Class. Quant. Grav.} \textbf{19} 2717-2741\\
    Bojowald M 2003 Homogeneous loop quantum cosmology
    \textit{Class. Quant. Grav.} \textbf{20} 2595-2615\\
    Ashtekar A, Bojowald M and Lewandowski J 2003
    Mathematical structure of loop quantum cosmology
    \textit{Adv. Theor. Math. Phys.} \textbf{7} 233-268
 \bibitem{predictions} Tsujikawa S, Singh P and Maartens R 2004
    Loop quantum gravity effects on inflation and the CMB
    \textit{Class. Quant. Grav.} \textbf{21} 5767-5775
 \bibitem{quantumbh_paps}
    Bojowald M 2004 Spherically symmetric quantum geometry: States and basic
      operators \textit{Class. Quant. Grav.} \textbf{21} 3733-3753\\
    Bojowald M 2004 The volume operator in spherically symmetric quantum
      geometry \textit{Class. Quant. Grav.} \textbf{21} 4881-4900\\
    Bojowald M and Swiderski R 2005
      Spherically symmetric quantum horizons \textit{Phys. Rev.} D
      \textbf{71} 081501\\
    Bojowald M, Goswami R, Maartens R and Singh P 2005
      A black hole mass threshold from non-singular quantum
      gravitational collapse \textit{Phys. Rev. Lett.} \textbf{95}
      091302\\
    Ashtekar A and Bojowald M 2005 Black hole evaporation: A paradigm
    \textit{Class. Quant. Grav.} \textbf{22} 3349-3362\\
    Ashtekar A and Bojowald M 2006 Quantum geometry and the Schwarzschild
    singularity \textit{Class. Quantum Grav.} \textbf{23} 391-411
 \bibitem{qft_refs1}
    Corichi A, Cortez J and Quevedo H 2004 On the relation between
    Fock and Sch\"{o}dinger representations for a scalar field
    \textit{Annals Phys.} \textbf{313} 446-478 (\textit{Preprint}
    \texttt{hep-th/0202070})\\
    Ashtekar A and Magnon A 1987 Quantum fields in curved Space-times
    \textit{Proc. Roy. Soc. Lond.} A \textbf{346} 375-394
 \bibitem{gj} Glimm J and Jaffe A 1987
    \textit{Quantum Physics: A Functional Integral Point of View}
    (New York: Springer-Verlag)
 \bibitem{rs} Reed M and Simon B 1980 \textit{Functional Analysis} vol. 1
    (San Diego: Academic Press)
 \bibitem{klauder} Klauder J 1999 Universal procedure for enforcing
    quantum constraints \textit{Nucl. Phys.} B \textbf{547} 397-412
    (especially pp 399-400)
 \bibitem{ashtekar1} Ashtekar A 1991 \textit{Lectures on Non-perturbative
    Canonical Gravity} (Singapore: World Scientific), appendix B
 \bibitem{tate} Tate R 1992 ``An algebraic approach to the quantization of
    constrained systems: finite dimensional examples''
    Ph.D. Thesis, Syracuse University, pp 111-115
 \bibitem{symmstates_paps}
    Bojowald M and Kastrup H A 2000
    Quantum symmetry reduction for diffeomorphism invariant theories
    of connections \textit{Class. Quant. Grav.} \textbf{17}
    3009-3043\\
    Bojowald M and Kastrup H A 2001
    Symmetric states in quantum geometry
    \textit{Preprint} \texttt{gr-qc/0101061}
 \bibitem{bt} Brunnemann J and Thiemann T 2006 On (cosmological) singularity
    avoidance in loop quantum gravity \textit{Class. Quantum Grav.} 
    \textbf{23} 1395-1427
 \bibitem{bs} Bojowald M and Strobl T 2000 Group theoretical quantization
    and the example of a phase space $S^1 \times \R^+$
    \textit{J. Math. Phys.} \textbf{41} 2537-2567
 \bibitem{compl_coh}
    Thiemann T 2006 Complexifier Coherent States for
    quantum general relativity \textit{Class. Quantum Grav.}
    \textbf{23} 2063-2117
 \bibitem{al} Ashtekar A and Lewandowski J 2001 Relation between
    polymer and Fock excitations \textit{Class. Quant. Grav.}
    \textbf{18} L117-L128
 \bibitem{conrady} Conrady F 2005 Free vacuum for loop quantum gravity
    \textit{Class. Quant. Grav.} \textbf{22} 3261-3293
 \bibitem{klauder_coh1}
  Klauder J 1962 Continuous-representation theory I. Postulates of
    continuous-representation theory
    \textit{J. Math. Phys.} \textbf{4} 1055-1058 \\
  Klauder J 1962 Continuous-representation theory II. Generalized
    relation between quantum and classical dynamics \textit{J. Math.
    Phys.} \textbf{4} 1058-1073
 \bibitem{klauder_coh2}
  Klauder J 1996 Coherent states for the hydrogen atom
    \textit{J. Phys.} A \textbf{29} L293-L298 \\
  Crawford M 2000 Temporally stable coherent states in energy degenerate
    systems: The hydrogen atom \textit{Phys. Rev.} A \textbf{62} 012104-1 \\
  Klauder J 2001 (June) ``The current state of coherent states''
    Contribution to the 7th ICSSUR Conference (\textit{Preprint}
    \texttt{quant-ph/0110108})
 \bibitem{wb}
  Willis J 2004 ``On the low-energy ramifications and a mathematical
    extension of loop quantum gravity'' Ph.D. Thesis, Penn State
    University, pp 74-78 \\
  Bojowald M and Skirzewski A 2005 Effective equations of motion for
    quantum systems \textit{Preprint} \texttt{math-ph/0511043}
 \bibitem{lqc_phys}
  Noui K, Perez A and Vandersloot K 2005 On the physical Hilbert space
    of loop quantum cosmology
    \textit{Phys. Rev.} D \textbf{71} 044025 \\
  Ashtekar A, Pawlowski T and Singh P 2006 Quantum nature of the big-bang
    \textit{Preprint} \texttt{gr-qc/0602086}
 \bibitem{abc}  Ashtekar A, Bombelli L and Corichi A 2005
    Semiclassical states for constrained systems
    \textit{Phys. Rev.} D \textbf{72} 025008
 \bibitem{mastconstr}
    Thiemann T 2003 The Phoenix project: Master constraint programme for
      loop quantum gravity \textit{Preprint} \texttt{gr-qc/0305080}\\   
    Thiemann T 2005 Quantum spin dynamics VIII. The master constraint
      \textit{Preprint} \texttt{gr-qc/0510011}.
\end{thebibliography}
\end{document}